\documentclass[preprint,12pt]{elsarticle}
\pdfoutput=1
\usepackage{longtable}
\usepackage{graphicx}
\usepackage{amssymb}
\usepackage{amsthm}
\usepackage{gensymb}
\usepackage{natbib}
\usepackage{lineno}
\usepackage{mathtools}
\usepackage{bm}
\usepackage{todonotes}
\usepackage{soul}
\usepackage{setspace}
\usepackage{caption}
\usepackage{booktabs}
\usepackage{array}
\usepackage{multirow}
\usepackage{rotating}
\usepackage[margin=1.0in]{geometry}
\biboptions{semicolon,round,authoryear}
\newcommand{\beginsupplement}{
\setcounter{table}{0}
\setcounter{figure}{0}
\setcounter{page}{1}
\setcounter{section}{0}
\setcounter{equation}{0}
        \renewcommand{\thetable}{S\arabic{table}}%
        \renewcommand{\thefigure}{S\arabic{figure}}%
        \renewcommand{\thesection}{S\arabic{section}}%
        \renewcommand{\thepage}{S\arabic{page}}%
        \renewcommand{\theequation}{S\arabic{equation}}%
     }

\begin{document}
\begin{frontmatter}

\title{Statistical modeling of rates and trends in Holocene relative sea level}

\author[RutgersStats,RutgersEPS,RutgersEOAS]{Erica L. Ashe}
\author[UCD]{Niamh Cahill}
\author[BostonColl]{Carling Hay}
\author[Nanyang1]{Nicole Khan}
\author[Tufts]{Andrew Kemp}
\author[RI]{Simon Engelhart}
\author[Nanyang1,Nanyang2,DMCS]{Benjamin P. Horton}
\author[UCD]{Andrew Parnell}
\author[RutgersEPS,RutgersEOAS]{Robert E. Kopp}

\address[RutgersStats]{Department of Statistics and Biostatistics, Rutgers University, New Brunswick, NJ, United States}
\address[RutgersEPS]{Department of Earth \& Planetary Sciences, Rutgers University, New Brunswick, NJ, United States}
\address[RutgersEOAS]{Institute of Earth, Ocean \& Atmospheric Sciences, Rutgers University, New Brunswick, NJ, United States}
\address[UCD]{School of Mathematics and Statistics, University College Dublin, Dublin, Ireland}
\address[BostonColl]{Department of Earth and Planetary Sciences, Boston College,Chestnut Hill, MA, United States}
\address[Nanyang1]{Asian School of the Environment, Nanyang Technological University, Singapore}
\address[Tufts]{Department of Earth and Ocean Sciences, Tufts University, Medford, MA, United States}
\address[RI]{Department of Geosciences, University of Rhode Island, Kingston, RI, United States}
\address[Nanyang2]{Earth Observatory of Singapore, Nanyang Technological University, Singapore}
\address[DMCS]{Department of Marine and Coastal Sciences, Rutgers University, New Brunswick, NJ, United States}

\begin{abstract}

Characterizing the spatio-temporal variability of relative sea level (RSL) and estimating local, regional, and global RSL trends requires statistical analysis of RSL data.  Formal statistical treatments, needed to account for the spatially and temporally sparse distribution of data and for geochronological and elevational uncertainties, have advanced considerably over the last decade.  Time-series models have adopted more flexible and physically-informed specifications with more rigorous quantification of uncertainties.  Spatio-temporal models have evolved from simple regional averaging to frameworks that more richly represent the correlation structure of RSL across space and time.  More complex statistical approaches enable rigorous quantification of spatial and temporal variability, the combination of geographically disparate data, and the separation of the RSL field into various components associated with different driving processes. We review the range of statistical modeling and analysis choices used in the literature, reformulating them for ease of comparison in a common hierarchical statistical framework.  The hierarchical framework separates each model into different levels, clearly partitioning measurement and inferential uncertainty from process variability. Placing models in a hierarchical framework enables us to highlight both the similarities and differences among modeling and analysis choices. We illustrate the implications of some modeling and analysis choices currently used in the literature by comparing the results of their application to common datasets within a hierarchical framework.  In light of the complex patterns of spatial and temporal variability exhibited by RSL, we recommend non-parametric approaches for modeling temporal and spatio-temporal RSL.
\end{abstract}


\end{frontmatter}

\section{Introduction}
\label{S:Intro}
 
The instrumental record of change in relative sea level (RSL, the difference between sea-surface height and land-surface height) is short, with the oldest tide-gauge record (Amsterdam, The Netherlands) dating to the 18\textsuperscript{th} century (e.g., \citealp{VanVeen1945TG,Woodworth1999}). Modern, quality-controlled measurements from Northern Hemisphere sites are available beginning in the early-to-mid 19\textsuperscript{th} century, and globally from the mid 20\textsuperscript{th} century onward. However, the geographic distribution of observations remains skewed to the Northern Hemisphere \citep{PSMSL2017,Holgate2013JCR,Pugh1987HB}. RSL proxies are therefore required to infer RSL changes and the contribution of processes that operate over longer timescales \citep{Bloom1964Peat,Shennan1989HolCrust,Trnqvist2008Miss,Dutton2015SLR}. Whereas instrumental records are (near-)continuous with relatively small vertical uncertainty and negligible (minute- to hourly-resolution) temporal uncertainties, RSL proxy data exhibit sample-specific vertical (inferential and measurement) and temporal uncertainties (e.g., \citealp{Trnqvist2015Handbook,Woodroffe2015Holo,Hibbert2016Coral}).  Like the distribution of tide-gauge measurements, the distribution of RSL proxy data is sparse in time and space. 

Quantifying the rates and spatial patterns of RSL change, on timescales ranging from decades to millennia, therefore involves piecing together sparse and noisy instrumental and/or proxy data (e.g., \citealp{Kopp2009probabilistic,Kopp2016PNAS,Hay2015probabilistic,Piecuch2017JRG}). Statistical models allow RSL records to be filtered for quality assurance \citep{Dusterhus2016LIG} as well as fused in a consistent manner that allows rigorous quantification of multiple sources of uncertainty. Statistical models are needed to answer fundamental questions in sea-level research, such as quantifying rates of RSL change (e.g., \citealp{Cahill2015modeling,Khan2015Holocene}), assessing spatial variability of the extent and magnitude of high stands (e.g., \citealp{Kopp2009probabilistic,Khan2017Carib,VacchiprepRSLNEC}), identifying the global-mean sea-level (GMSL) signal (e.g., \citealp{Church2004Estim,Jevrejeva2006,Kopp2009probabilistic,Hay2015probabilistic,Kopp2016PNAS}), and improving estimates to constrain dominant physical processes, including ice-sheet behavior and glacio-isostatic adjustment (GIA; \citealp{Engelhart2011HolGIA,Mitrovica2011FP}), based on their distinct spatial and temporal patterns (e.g., \citealp{Milne2005HolCar,Dangendorf2017GMSL,Kopp2015GeogVar,Hay2015probabilistic}).  Although statistical methods have for decades played a major role in reconstructing other paleoclimate variables (e.g., temperature; \citealp{Visser1988KF,Fritts1991Book,Smith1996EOFSST,Mann1998Temp}), their application to instrumental (e.g., \citealp{Church2004Estim,Jevrejeva2006,Hay2013GSL,Kopp2013grl}) and paleo (e.g., \citealp{Parnell2005thesis,Kopp2009probabilistic,Cahill2016bayesian,Khan2015Holocene}) RSL data is more recent. 

Hierarchical statistical models, described in detail in Section \ref{S:Framework}, distinguish between a process level (representing, for example, the physics of RSL change) and a data level (representing, for example, the noisy recording of RSL by instruments or proxies) and cleanly distinguish between variability and uncertainties introduced at these different levels \citep{Cressie2015statistics, Tingley2012piecing}.  They are flexible, capable of accommodating missing data, and enable probabilistic inference about RSL over time and space. Viewing statistical models in a hierarchical framework, however, does not require a hierarchical computational implementation; the hierarchical perspective provides a valuable tool for dissecting and comparing models \citep{Tingley2010BARCAST}, regardless of implementation.  Though only some authors have used hierarchical RSL models explicitly (e.g. \citealp{Kopp2016PNAS,Khan2017Carib}), almost all statistical models of RSL can be reformulated hierarchically. Using a hierarchical framework to present \textit{modeling choices} (i.e., how to characterize the relationships among variables; Italicized terms are defined in the glossary, Section \ref{tab:Glossary}) and \textit{analysis choices} (i.e., how to implement a model structure) in a consistent manner, we present an integrated perspective on the choices made in analyzing temporal and spatio-temporal RSL datasets. Although this paper primarily concentrates on proxy data from the Holocene, the models are applicable to other timescales (e.g., \citealp{Kopp2009probabilistic}). The appropriate modeling and analysis choices depend on the research questions asked, the type of data used, and the spatio-temporal scale (e.g., local to global, years to millennia) under consideration. 

In the remainder of this paper, we first introduce hierarchical models, which consist of several levels, and their application to RSL data (Section \ref{S:Framework}). We then define different models representing the data-generating process (Section \ref{SS:data}) by which RSL is linked to proxy records. At the process level, we describe time-series models (Section \ref{S:TimeSeries}) for representing RSL at a single site and spatio-temporal models (Section \ref{S:STMods}) for representing the temporal evolution of sea level across a regional or global domain.  We then discuss different analysis techniques and their advantages and disadvantages (Section \ref{S:Analysis}). To illustrate more concretely the similarities and differences between these approaches, we build case studies by applying models to common datasets including tide-gauges measurements (TGs), near-continuous RSL reconstructions (\textit{continuous cores}), and \textit{sea-level index points (SLIPs)} from locations along the Atlantic coast of the United States (Section \ref{S:Implement}).  Finally, we make recommendations to help identify which methods to use to obtain temporal and spatio-temporal estimates of RSL and/or GMSL and rates of change based on the data being analyzed and the objective of the study (Section \ref{S:Conclusion}).  

\section{Hierarchical statistical modeling}
\label{S:Framework} 

In statistical nomenclature, \textit{uncertainty} signifies an interval around which the true value is likely to fall, whereas statistical \textit{error} is the (unknown) difference between the predicted value and the true value. \textit{Residuals}, which can be analyzed to test assumptions about modeled errors, are the difference between an observed and a predicted value (Section \ref{fig:Resid}). The \textit{prior distribution} of a Bayesian model represents the knowledge about a given phenomenon before new data is observed, whereas the \textit{posterior distribution} is the conditional probability that is assigned after the relevant new evidence (the observed data) is taken into account.

Hierarchical statistical models, which are frequently but not always implemented in a Bayesian framework, partition the multiple random effects that lead to individual observations into levels, thus clarifying the assumptions in a statistical analysis. They separate the underlying phenomenon of interest, such as sea level, and its variability, characterized at what is called the \textit{process level}, from the noisy mechanism by which this underlying process is observed, characterized at the \textit{data level}).  Bayesian hierarchical models are based on conditional probabilities: observed data are regarded as conditional on a \textit{latent} (unobserved) process, which is conditioned on unknown parameters and the assumptions in the model structure.  Inverting the conditional probabilities allows probabilistic estimation of a time series or field, which can vary as a function of time and/or space. Each level of a hierarchical statistical model quantifies the uncertainties of that level separately; this can require more careful consideration of sources of uncertainties than approaches that pool the uncertainties from different levels together. Almost any statistical model can be reinterpreted as a hierarchical model; doing so increases transparency by explicitly making the distinction between modeling assumptions and analysis methods (or inference choices), as well as the difference between process variability and observation \textit{noise}.

The primary goal in statistical analysis of RSL data is to estimate latent RSL (i.e., the noise-free time series or spatio-temporal field) and its uncertainty from observed, \textit{noisy data}. At least three levels are defined in most RSL model hierarchies.  The data level characterizes the relationship between RSL and the observed RSL data (instrumental and/or proxy) and incorporates measurement, inferential (e.g., from the conversion of a proxy's elevation to a distribution of likely RSL), and dating uncertainties. The process level models `true' (i.e., noise-free) RSL and in some cases, decomposes RSL into the underlying processes that comprise it. The parameter level captures key attributes of the data and process levels through unobserved parameters (e.g., characteristic temporal and spatial scales of variability).  Hierarchical models estimate the posterior probability distribution of the noise-free RSL time series or field (and its uncertainty), which enables probabilistic inference about RSL over time and space (see \citealp{Cressie2015statistics}, for further details on hierarchical models). We interpret published sea-level analyses within a hierarchical framework in order to compare modeling assumptions as well as analysis methods from these implementations. 
\label{SS:Framework}
Conditional probability distributions are the basic mechanism for modeling uncertainty in hierarchical models. The \textit{conditional probability} distribution of A, given B, is denoted $p(A|B)$.  Bayesian statistics uses \textit{Bayes' theorem} \citep{Laplace1812Bayes} to invert the conditional probability of the observed data, $y$, and calculate the conditional probability of unknown parameter(s) or process(es), $\theta$, given the data, $y$:

\begin{equation}
\label{eq:Bayes1}
p(\theta|y) = \frac{p(y|\theta) p(\theta)}{p(y)}.
\end{equation}
\noindent The \textit{likelihood} function, $p(y|\theta)$ (also known as a sampling or data distribution), is the probability of observing the data as described by the parameter(s) or process(es) of the fitted model.  The prior distribution, $p(\theta)$, expresses a priori beliefs about the unknown parameter(s) or process(es), before data have been observed, and $p(y)$ is the \textit{marginal} likelihood of the data, defined as the probability of observing $y$ averaged across all possible processes or parameters.  The conditional, posterior distribution, $p(\theta|y)$, is the resulting process or parameter distribution, given the observations. The parameters used to construct the prior distribution, known as \textit{hyperparameters}, can be fixed, estimated, or have \textit{(hyper)prior} distributions themselves. For the remainder of this paper, we will ignore the marginal likelihood, which is irrelevant provided the observations are static, and use the alternative form of Bayes' theorem that states the posterior is proportional to the likelihood times the prior:

\begin{equation}
\label{eq:Bayes2}
p(\theta|y) \propto p(y|\theta) p(\theta).
\end{equation}
In a simple hierarchical statistical model of RSL, the data model, $p(y|f,\theta_d)$, expresses the distribution of the RSL data, $y$, given the latent (unobserved) sea-level process, $f$, and the parameters of that distribution, $\theta_d$.  Below the data level, the RSL process model, $p(f|\theta_s)$, incorporates scientific knowledge and uncertainty into the estimation of the true RSL process through its conditional parameters, $\theta_s$.  On the bottom level, the parameter model, $p(\theta_d,\theta_s)$, specifies the prior distribution of all unknown parameters and hyperparameters.

\begin{equation}
\underbrace{p(f,\theta_s,\theta_d|y)}_{\text{posterior}}\propto \underbrace{p(y|f,\theta_d)}_{\text{data model}}\cdot \underbrace{p(f|\theta_s)}_{\text{process model}}\cdot \underbrace{p(\theta_d,\theta_s)}_{\text{parameter model}}
\end{equation}

\label{SS:AnModChoice}
Modeling choices refer to the relationships defined within a model and the assumptions made in constructing these relationships (e.g., a linear relationship between time and RSL), whereas analysis choices describe decisions about how to implement a specific model structure (e.g., using least-squares analysis, \citealp{Aitken1934GLS}; likelihood maximization, \citealp{Wilks1938LikelihoodRatio}; or fully Bayesian analysis with Monte Carlo sampling, \citealp{Hastings1970MCSampling}).

Models are always simplified, imperfect versions of reality.  It is therefore important to recognize that a model's estimate of the truth (e.g., the latent process, $f$) is conditional upon the assumptions of the model and the accuracy of the analysis approach.  Because a perfect model of the world is also a uselessly intractable one \citep{Borges1975}, statistical estimates are useful, but imperfect, approximations.  Consideration of alternative sets of structural modeling assumptions is an important part of characterizing the robustness of an estimate.

The hierarchical statistical framework accommodates a broad range of complexity in modeling and analysis choices, and most methods of statistical analysis used in sea-level science can be reframed as hierarchical models. For example, the structure of trends in RSL through time can be defined prior to analysis by explicitly assuming linear, polynomial, piecewise-linear, or other forms of the relationship between time and RSL at the process level. \textit{Non-parametric} approaches, such as spline regression (e.g., \citealp{Gharineiat2015splines}) or models with \textit{Gaussian process} priors (GP; \citealp{Rasmussen2006GP}; e.g., \citealp{Kopp2009probabilistic,Cahill2015modeling}), can also be used to determine trends, without a predetermined functional form at the process level.  A probabilistic ensemble approach (e.g., \citealp{Dusterhus2016LIG}), where each ensemble member is assigned an equal prior probability, is another option for modeling the process level. Table \ref{tab:Models} presents pairs of modeling and analysis choices from past analyses, recast within a hierarchical framework, highlighting the time periods and data that have been analyzed in the literature using these models.

\begin{table}[h!]
  \tiny
  \caption{Techniques table}
 \label{tab:Models} 
 \begin{tabular}[l]{llllll}
\textbf{Technique} &\textbf{Analysis Methods} &\textbf{Modeling Choices} &\textbf{Data} &\textbf{Time Period} &\textbf{Examples in Publications}\\ [0.5ex]
  \midrule
Simple Linear 	&\multirow{2}{*}{Least squares}&\multirow{2}{*}{Temporally linear	}&\multirow{2}{*}{TGs,CCs, SLIPs}	&\multirow{2}{*}{$\leq 3$ ky}	&Shennan et al. (2002)\\Regression&&&&& Engelhart et al. (2009)\\
 \midrule
\multirow{2}{*}{EIV Change-point}&Errors-in-variable,	&\multirow{2}{*}{Change-point model}&\multirow{2}{*}{CCs,TGs, SLIPs}	&Common Era,	&Kemp et al. (2013), \\
&Bayesian analysis&&&Late Holocene&Brain et al. (2015)\\
 \midrule
\multirow{3}{*}{EIV IGP}	&\multirow{2}{*}{Errors-in-variable,} &Covariance functions, 	&\multirow{3}{*}{TGs,CCs,SLIPs}	&\multirow{2}{*}{Common Era,}&Gehrels et al. (2013), \\
&\multirow{2}{*}{Bayesian analysis}	&{Proxy-systems model,}&& \multirow{2}{*}{Holocene}&Cahill et al. (2015)\\
&& Integrated GP\\
 \midrule
\multirow{3}{*}{Regional Averaging}&Least squares, &\multirow{3}{*}{Physical models}	&\multirow{3}{*}{TGs, Altimetry data}	&\multirow{3}{*}{Instrumental}&Douglas et al. (1991),\\
& Ad hoc,  &&&&Jevrejeva et al. (2009),\\
&Virtual station&&&&Dangendorf et al. (2017)\\
 \midrule
\multirow{2}{*}{EOF Regression}	&\multirow{2}{*}{Least squares}	&\multirow{2}{*}{EOFs}&\multirow{2}{*}{TGs, Altimetry data}	&\multirow{2}{*}{Instrumental}&{Church \& White (2006)}\\
&&&&&Church et al. (2004, 2011)\\
 \midrule
Probabilistic&\multirow{2}{*}{Particle filter}	&\multirow{2}{*}{Physical models}	&	\multirow{2}{*}{SLIPs}&\multirow{2}{*}{LIG}&\multirow{2}{*}{D\"usterhaus et al. (2016)}\\
Ensembles&&&&&\\
 \midrule
\multirow{3}{*}{Kalman Smoother}	&\multirow{3}{*}{Multi-model KS}	&Spatio-temporal,	&\multirow{3}{*}{TGs}	&\multirow{3}{*}{Instrumental} &\multirow{2}{*}{Hay et al. (2013,} \\&&Covariance functions,&&&\multirow{2}{*}{2015, 2017)}\\
&&Physical models\\
 \midrule
\multirow{3}{*}{Gaussian processes}	&{Bayesian Analysis,} &Spatio-temporal, &\multirow{3}{*}{TGs,CCs, SLIPs}&Instrumental, 	&Parnell (2005), Kopp (2013) \\
&Empirical Bayesian	&Covariance functions&&Holocene, LIG,&Kopp et al. (2015), \\
&analysis&Physical models&&Common Era&Khan et al. (2015, 2017)\\
 \midrule
\end{tabular}
Table includes common techniques, analysis methods, modeling choices, the type of data typically used, relevant time periods to which this approach has been applied, and some examples in publications.  Sections \ref{S:TimeSeries} and \ref{S:STMods} provide details on the modeling choices, and section \ref{S:Analysis} discusses specific analysis choices. TGs - tide gauges; CCs - continuous core records; SLIPs - sea-level index points; EIV - errors-in-variables; IGP - integrated Gaussian process; EOFs - empirical orthogonal functions.
\end{table}

\section{Modeling Choices}
\subsection{Modeling the data level}
\label{SS:data}

RSL proxy data differ from instrumental data in their sources of uncertainty, which are modeled at the data level.  Whereas instrumental data have negligible temporal uncertainty, proxy data have inherent temporal uncertainties (e.g., from radiometric dating; \citealp{Polach1976Rad,Stuiver197714C,Reimer2013}). In the broadest sense, there are two types of Holocene RSL proxy data. Discrete SLIPs constrain the position of RSL in time and space (geographic and vertical) and can be treated independently of one another at the data level in most circumstances (e.g., \citealp{Shennan2002Comp,Engelhart2012QSR,Engelhart2015Pac}).  In contrast, continuous cores are produced by analyzing a sequence of ordered samples from a single sediment core.  These records constrain RSL change through time at a single geographic location \citep{Gehrels2002,Varekamp1992TN}, but the data points are not independent because a common age-depth model (e.g., \citealp{Wright2017AgeDepth}) is generally used to estimate sample age. A particular challenge when working with RSL proxy data is realistically characterizing the geochronological uncertainties that arise from the process of radiocarbon calibration \citep{Reimer2013}, which results in probability distributions for calendar ages that are often multi-modal and discontinuous.  However, many models assume (explicitly or implicitly) normal uncertainties for calibrated radiocarbon ages for simplicity (although this is an oversimplification of reality). Some age-depth models (e.g., \citealp{Parnell2008Bchron,Parnell2015SVM,Wright2017AgeDepth}) handle these difficulties and return predicted age distributions with approximately normal uncertainties.

\label{SS:data_level}
The data level of a hierarchical statistical model represents the relationship between uncertain observations and RSL. The specific type of data and associated uncertainty determine the form of this relationship and hence the form of the data-level model.  For example, regression models often assume that the independent variable time, $t$, has been measured exactly, and only account for uncertainty in time's functional relationship with RSL, $f$.  This link between the observed data and the sea-level process can be represented as

\begin{equation}
\label{eq:TSData}
y_i = f(t_i) + \varepsilon_i,
\end{equation}

\noindent where $y_i$ is proxy or instrumental observation $i$ and $f(t_i)$ is true RSL (under the assumptions of the model) at the time that $y_i$ was observed.  Many models assume measurement uncertainties are independent and normally distributed, such that $\varepsilon_i \sim \mathcal{N}(0,\sigma_i^2)$, where $\sigma_i$ is the assumed standard deviation of measurement and \textit{inferential uncertainty} for observation $i$ (e.g., \citealp{Hijma2015Handbook}).  This assumption, however, typically ignores some biases (e.g., miscalibration) and assumes an ideal measurement is taken, which is rarely achieved in reality.  In analyses that do not incorporate measurement and inferential uncertainty specific to each observation, $\varepsilon$ (equation \ref{eq:TSData}) is typically assumed to be independent and identically distributed (iid) Gaussian uncertainty and pools data uncertainty with process variability not represented in the structure of $f(t)$. The data level of a spatio-temporal model is equivalent to that of a time-series model, where true sea-level, $f(\bm{x}_i,t_i)$, is dependent on both geographic location, $\bm{x}_i$, and time, $t_i$. 

The distinction between measurement uncertainty and inferential uncertainty (the relationship between RSL and a proxy's position) can be explicit:

\begin{equation}
\label{eq:dat_lev}
 y_i = f(t_i) + \varepsilon_i + \eta_i
\end{equation}

\noindent where $\varepsilon_i$ is the unobserved measurement error for observation $i$, and $\eta_i$ is the indicative meaning (vertical relationship of a proxy to contemporary tide levels) uncertainty for the specific sample ($\eta=0$ for direct instrumental observations or RSL), which may depend on time due to changes in tidal range. Hierarchical models for RSL proxies can be even more explicit in the representation of $\eta_{i}$; whereas this is often specified in a database based upon an interpretation (which can introduce subjectivity and additional assumptions) conducted separately, it can be related directly and probabilistically to raw data, such as microfossil species abundances, in an additional level of the model (e.g., extending the approach used by \citealp{Parnell2015SVM} for paleoclimate data). 

Temporal uncertainties in proxy data are separated from process noise at the data level:

\begin{equation}
\label{eq:observt}
\overline{t_i} = t_i + \delta^t_i, 
\end{equation}

\noindent where $\overline{t_i}$ is the central point estimate of the calibrated age for radiocarbon dating, $t_i$ is the true age (under the assumptions of the model), which is unknown and unobserved, and $\delta^t_i$ is unobserved temporal error, which is often incorporated as normal uncertainty within the analysis. These uncertainties can be incorporated in several different ways (see Section \ref{SS:Temp_Imp}).

\subsection{Process level: Modeling the temporal sea-level process}
\label{S:TimeSeries}

RSL time-series models have a long history, beginning with hand-drawn curves (e.g., \citealp{Lighty1982Acrop,Zong2004Hand-Drawn,Smith2011Hand-Drawn,Abdul2016YoungerDryas}) and evolving to include different forms of statistical regression (e.g., temporally linear, \citealp{Shennan2002Comp}; change-point, \citealp{Kemp2015Paleo}; GP, \citealp{Kopp2009probabilistic}; EOF, \citealp{Church2004Estim}). Some of these explicitly separate data uncertainty from process variability; others incorporate both data uncertainty and non-linear or high-frequency process variability into the error term, $\varepsilon$.  Recasting these models in a hierarchical framework allows the separation of uncertainties of different types, providing a common basis for comparing modeling choices.  

\subsubsection{Temporally linear models}
\label{SS:LR}

The simplest approach to estimating RSL and an average rate of RSL change is fitting a temporally linear model to observed data. As just two examples, \cite{Shennan2002GB} and \cite{Engelhart2009Geol} applied simple linear regression to discrete SLIPs and tide-gauge measurements to estimate the rate of RSL change during the past few thousand years, over which period the observations were qualitatively judged to be well approximated by a linear trend.  In both instances, the authors performed linear regression on the midpoints of the SLIPs and did not account for inferential and measurement uncertainty (temporal and vertical).  The process-level relationship is represented by

\begin{equation}
\label{eq:LinReg}
f(t) = m\cdot t + \beta,
\end{equation}

\noindent where $f(t)$ is the modeled true RSL, $m$ is the constant rate of change in RSL, and $\beta$ is the intercept. The slope, $m$, and y-intercept parameter, $\beta$, can be estimated using many analysis methods, but are most typically analyzed using least-squares regression.  

Linear models are familiar to most researchers, easy to use, and therefore provide a convenient way to find a first-order estimate (a rough approximation) of rates over time periods when they are expected to be roughly constant.  However, linear models can provide biased estimates of the slope parameters, due to their sensitivity to the temporal distribution of data.  For example, intervals with a high concentration of data exert an undue influence on rate estimates. In addition, the linearity assumption is rigid; linear models lack the ability to model any evolution in rates of RSL change. Linear regression also assumes stationarity of errors (errors do not change over time or among observations) when using ordinary least squares regression.  Linear models are appropriate for modeling a first-order estimate, but are generally inappropriate for more in-depth analyses.

\subsubsection{Change-point models}
\label{SS:cpr}

Change-point models represent a single time series as separate, continuous, temporally linear sections and are generally employed to estimate the timing of changes in trend.  For example, \cite{Kemp2015RSLConn} estimated when modern rates of RSL change began in Connecticut using change-point models.  \cite{Long2014CP} used a change-point model to analyze whether there was an acceleration in RSL change in the UK over the past 300 years.  At the process level, with $m$ change points,

\begin{equation}
\label{eq:CPR}
f = 
\begin{cases}
\alpha_1 + \beta_1(t - \gamma_1), & \text{when}\ t < \gamma_1\\
\alpha_1 + \beta_2(t - \gamma_1), & \text{when}\ \gamma_1 < t < \gamma_2\\
\alpha_{j-1} + \beta_j(t - \gamma_{j-1}), &\text{for}\ j = 3,\ldots,(m+1),\ \text{and}\ t>\gamma_{j-1},
\end{cases}
\end{equation}
\noindent where $\gamma_k$ is the change point and $\alpha_k$ is the expected value of RSL at the change point (with a continuity constraint, such that $\alpha_{k}=\alpha_{k-1}+\beta_{k-1}(\gamma_{k}-\gamma_{k-1})$), and $\beta_j$ is the rate of RSL change for each of the $m+1$ segments. The parameters of change-point models can be estimated using a range of analysis approaches, including non-linear least squares and empirical Bayes (Section \ref{SS:EHM}), but in the RSL modeling literature, these models generally follow \cite{Cahill2015CP} in employing a change-point process model using a \textit{Bayesian Hierarchical Model} (BHM; Section \ref{SS:BHM}) within an \textit{errors-in-variable (EIV)} framework (Section \ref{SS:EIV}).

Change-point models attempt to address a primary goal in sea-level research, identifying accelerations in RSL and GMSL change (e.g., \citealp{Church2006Accel,jevrejeva2008recentaccel,Kopp2013grl}), and they improve upon simple linear models by allowing for varying rates of RSL change and are relatively simple to implement.  However, the linear constraints on each section are still fairly rigid and often do not represent the true physical behavior of RSL. When there is a clear pattern of phase changes in the data and variability around the trends is \textit{white noise} (signal having serially uncorrelated random variation), change-point models may be appropriate for estimating the timing of these phase changes; however, they cannot estimate the magnitude of accelerations because they assume acceleration is instantaneous.  Additionally, the white noise assumption, which can be tested with analyses of residuals, is frequently violated.  If the model accounts for every change point in the regression lines, the assumption of iid errors can be met.  Alternative, less parametric approaches, such as \textit{Kalman Smoother} (KS) or GP models, are more flexible in representing RSL time series when the data exhibit fluctuations that cannot be adequately captured by a small number of change points; as such, they can help to answer questions about accelerations in a manner that recognizes that accelerations may occur gradually rather than abruptly.  For a more complete overview of change-point models, see \cite{Ducre2003ChangePoint}.

\subsubsection{Gaussian process models}
\label{SS:GP}
A Gaussian process (GP) is a generalization of the Gaussian (normal) probability distribution in continuous time (and space) \citep{Rasmussen2006GP}.  To our knowledge, GP modeling was introduced into sea-level analysis by \cite{Parnell2005thesis} and into the paleo-sea-level literature by the Last Interglacial analysis of \cite{Kopp2009probabilistic}. In a GP, the relationship among any arbitrary set of points (e.g., in time, or in space and time) is a multivariate normal distribution defined by a mean vector and a covariance matrix.  A temporal GP is fully defined by its mean function, $\mu(t)$, and \textit{covariance function}, $K(t,t')$, where $t$ is an input variable, which here represents time (but can be extended without loss of generality to higher dimensions, for example to include geographic location in spatial sea-level modeling; see Section \ref{S:STMods}).  When RSL, $f(t)$, is a GP this is expressed as

\begin{equation}
\label{eq:GP}
f(t)\sim \mathcal{GP}(\mu(t),K(t,t')).
\end{equation}
\noindent The covariance function, $K(t,t')$, defines prior expectations about the variance of the process about its mean and the correlation between points in time (and space), and thus about the way in which information is shared between time points. 

In a GP model, the sea-level function, $f(t)$, is non-parametric (i.e., its form is not predetermined). Accordingly, GP time-series models have much more flexibility than temporally linear or change-point models.  The shape of the curve is driven by the covariance matrix, which is estimated through the data, as opposed to a predetermined functional form. A key assumption of a GP model is that predictions at any given point assume a normal distribution.

\label{SS:IGP}
A variant type of GP model is an Integrated Gaussian Process (IGP) model, which places a GP prior on the rate process rather than the sea-level process (see \citealp{Holsclaw2013GPDerivCurves} for more details and justification).  The IGP approach was introduced into the RSL literature by \cite{Cahill2015modeling}, following the methodology of \cite{Holsclaw2013GPDerivCurves}, who presented a new method of obtaining the derivative process by viewing this procedure as an inverse problem.  At the process level, IGP regression models the RSL rate process, $f'(t)$,  as a GP.  The underlying RSL process, $f(t)$, is the integral of the rate process plus a constant intercept, $\alpha$:

\begin{align}
\label{eq:IGP}
f'(t)\sim\mathcal{GP}(\mu(t),K(t,t')),\\
f(t) = \alpha+ \int_0^{t} f'(u) du, 
\end{align}

\noindent where $t$ is true time. For example, \cite{Cahill2015modeling} estimated the continuous and dynamic evolution of RSL change in North Carolina from sediment cores using IGP models.
 
Assuming a stationary covariance for the rates of RSL change produces a non-stationary covariance for RSL.  The Bayesian approach allows regularization (introducing additional information in order to prevent over-fitting or solving an ill-fit problem), which reduces issues with identifiability (the theoretical possibility of learning the true values of a model's underlying parameters after obtaining an infinite number of observations from it). One limitation of the IGP is that the sea-level function needs to be two times differentiable; unlike a GP model of levels, this does not allow abrupt changes of rate, but instead requires that any change of rate happens through a gradual acceleration.  For example, any RSL trend that is well-represented by a change-point model, which assumes instantaneous acceleration, cannot be represented well by an IGP model.  Another drawback of the IGP is that it is not immediately clear how to extend it to define a derivative process in multiple dimensions (e.g., \citealp{Holsclaw2013GPDerivCurves}), such as for applying it to spatial datasets, although some insights could be gleaned from spatial first-difference methods, commonly applied in the econometrics literature (e.g., \citealp{Juodis2018FD}).

In both GP and IGP models, the covariance functions can take a range of functional forms (Section \ref{sup:cov}). The form and parameters of the covariance function -- called hyperparameters, because they set assumptions that inform the non-parametric representation of $f(t)$ -- define how abruptly modeled RSL may change with temporal (or spatio-temporal) distance. Scale hyperparameters express prior beliefs about the amplitude of variability over time. Range hyperparameters (or characteristic length scale) set the distance over which the correlation between two sites or times decays toward zero (e.g., \citealp{Rasmussen2006GP}). Smoothness parameters determine the speed of decay in the correlation in time or space (e.g., the degree of differentiability). For fixed hyperparameters, GP posterior distributions are analytically tractable (i.e., no approximation or sampling is necessary) when data uncertainty is represented as normally distributed; statistically speaking, this reflects the fact that the normal distribution is self-conjugate. Covariance functions for GP priors can be constructed by summing different terms with different characteristic scales of variability; however, linking these different terms with distinct physical processes requires incorporating process knowledge through deterministic physical models.

GP and IGP models are appropriate for many applications because of their flexibility and ability to incorporate physical knowledge regarding ranges and scales of variability through their covariance functions (Section \ref{sup:cov}).  However, they do have several key disadvantages. GP models generally assume that the covariance function is stationary -- e.g., that prior expectations about the relationship between RSL at 10 ka and 8 ka are the same as those between 4 ka and 2 ka. This is a rough approximation, although still more flexible than parametric approaches.  IGP models generally make the same assumption about rates as opposed to levels, which is a potentially more accurate approximation.  

GP models are considerably more difficult to implement than linear or change-point models, although an increasing number of tools are available to assist in their implementation (e.g., \cite{Kopp2016CESLCS}; \citealp{Cahill2016IGP}).  \cite{Kopp2016CESLCS} makes documentation and code publicly available for implementing a spatio-temporal version of a GP model, and \cite{Cahill2016IGP} provides code to implement an IGP model.  

These models exhibit relatively long analysis times (e.g., see Table \ref{t:mods} for analysis times of illustrative analyses). Statistical modeling is an iterative process of model development, model fitting, and model criticism, and slow analysis methods can be a hindrance to this process. Moreover, for some methods the computational time can scale rapidly with the number of data points. The time to invert a covariance matrix for a GP analysis scales with the cube of the number of data points, and the computational time of a model that both inverts a covariance matrix and samples temporal uncertainty (e.g., the EIV-IGP) scales with the number of data points to the fourth power. Although there are techniques to estimate the covariance matrix in order to make it more easily invertible, these models are currently not scalable to large datasets using full covariance matrices.

\subsubsection{Summary of time series models}
Each modeling choice has advantages and disadvantages. Temporally linear models are sensitive to the temporal distribution of data and influential data points. However, when uncertainties in the data are incorporated into the model, linear regression provides an easy, fast, and appropriate way to determine first-order rates of change in processes that are approximately constant. Change-point models assume that phases of persistent sea-level behavior are approximated by linear trends, which may not accurately represent the underlying physics of RSL change and mask (to some degree) the continuous evolution of RSL through time \citep{Cahill2015modeling}. Drawbacks of these simpler approaches motivated \cite{Cahill2015modeling} to develop a non-parametric (EIV-IGP) methodology for estimating rates of RSL change from multiple types of proxy data at a single site (Sections \ref{SS:CE} and \ref{SS:CompHol}) and \cite{Kopp2009probabilistic} and \cite{Kopp2013grl} to develop spatio-temporal GP models (Sections \ref{SS:CompHol} and \ref{SS:Comp1900}), which share information from nearby geographical sites to overcome the limited length of records in certain locations. However, GP models also have several drawbacks, including their less intuitive nature, complexity of implementation, longer computation time, and stationarity assumptions.   

\subsection{Process level: Modeling the spatio-temporal sea-level process}
\label{S:STMods}\label{SS:Spatial}
Spatio-temporal models allow information to be shared among sites based on their proximity or physical relations, and they also enable estimation of RSL and its rates of change at sites where there are no data. In addition, spatio-temporal models support the estimation of multi-site metrics, such as change in global-mean sea level (GMSL), which is defined as the spatial average of RSL or SSH (sea-surface height) over the ocean \citep{Gornitz1982GMSL}. Most spatio-temporal models implemented in the literature are not explicitly hierarchical, but -- as with time-series models -- they can be recast in this way in order to facilitate comparison.

Spatio-temporal RSL models represent a continuum from purely statistical to purely physical models. At the purely statistical end of this spectrum, the priors of the process level relating RSLs from different locations to one another are based solely on their spatial and temporal proximity, and the bounds on the hyperparameters (before they are optimized) are typically based on knowledge of the variability of the processes they attempt to capture. At the purely physical end of the spectrum, a deterministic model (e.g., a GIA model) is used to estimate the RSL field; probabilistic ensembles are just one example. Intermediate formulations incorporate physical information into the construction of prior distributions.

\subsubsection{RSL represented with single or multiple GP priors}

The simplest and most fully statistical models place a single GP prior on RSL, as in equation \ref{eq:GP} (with the mean and covariance functions dependent on both time and geographic location), conditioning on RSL proxy or instrumental data to yield a posterior distribution of RSL in time and space.  The covariance function in this context may be spatially and temporally separable, in which case it is represented as the product of a temporal covariance function and a spatial covariance function. The former describes prior expectations about scales of change in time, the latter about scales of change in space. The analysis of a spatio-temporal GP is amenable to the same approaches as a temporal GP.

A single GP with a parametric covariance function is rarely implemented in the spatio-temporal RSL modeling literature, because a single scale of temporal variability and a single scale of spatial variability is too simple to capture physical behavior. More physical insight recognizes that RSL should have multiple spatio-temporal scales of variability, and can therefore be represented as the sum of multiple terms with GP priors.  \cite{Kopp2013grl} introduced this approach into the spatio-temporal RSL literature to model tide-gauge data along the east coast of the United States in order to determine whether there was an acceleration in local RSL. His process model employed nine separate terms with GP priors, combining three spatial scales of variability (global, regional, and local) with three temporal scales of variability (low, medium, and high frequency). Lower resolution RSL proxy data frequently require a simpler process level. For example, several studies (e.g., \citealp{Kopp2016PNAS,Khan2017Carib}) employ models of the form:

\begin{equation}
f(\bm{x},t)=g(t) +r(\bm{x},t)+m(\bm{x},t)+l(\bm{x},t),
\label{eq:pure_stat}
\end{equation}

\noindent where $g(t)$ represents a global term that is common to all sites and could include (in a global analysis) the global-mean effects of thermal expansion and changing land ice volume; $r(\bm{x},t)$ is a regional term, which might represent processes like GIA, ocean/atmosphere dynamics, and the static-equilibrium effects of land-ice mass changes; $m(\bm{x},t)$ might capture smaller-scale region (or local) processes, like tectonics or natural sediment compaction; and $l(\bm{x},t)$ captures site-specific linear signals.

Although covariance functions can be used to capture specific physical processes that influence RSL, they have fundamental limitations.  Using covariance functions alone, based on scales of variability, can fail to represent the processes intended, in contrast to using physical models. For example, many common covariance functions will miss spatial teleconnections such as those associated with sea-level fingerprints or large-scale climate modes, because they assume that correlation decays with distance.  \cite{Hay2015probabilistic} used GPs but worked around this by using physical models to estimate the covariance, as did \cite{Kopp2009probabilistic}. However, the assumption that the distribution of physically probable outcomes is normal is restrictive, so this approach is also imperfect. 

\subsubsection{Empirical orthogonal functions}
\label{SS:emp_orth}
\textit{Empirical orthogonal function} (EOF) regression fits sparse observations with a set of spatial patterns (EOFs) that characterize the maximum amount of variation in a relatively dense, complementary dataset.  These spatial patterns are derived through EOF decomposition (equivalent to principal component analysis [PCA]), which decomposes the dense dataset into orthogonal patterns.  For example, \cite{Church2004Estim}, \cite{Domingues2008EOF}, and \cite{Ray2011Exp} used EOF decomposition to calculate the dominant spatial patterns of (high-frequency) variability in GIA-corrected SSH from altimetry observations, and applied these patterns to fit tide-gauge data and estimate longer-term GMSL change. Although EOF decomposition incorporates no direct or explicit knowledge of physical processes, many of the dominant EOFs are associated with known physical modes of variability (e.g., North Atlantic Oscillation, Atlantic Multi-decadal Oscillation, Pacific Decadal Oscillation, El Ni\~no Southern Oscillation). The process level in EOF regression can be represented as:

\begin{equation}
\label{eq:EOF}
f(\bm{x},t) = g(t) + \Sigma_i{U_i(\bm{x})\alpha_i(t)} + GIA(\bm{x},t-t_0).
\end{equation}
\noindent 
Here, $g(t)$ is a global `mode' that is constant over space, each $U$ represents a leading spatial EOF, $\alpha$ is a time series of amplitudes of the EOFs (also known as a principal component of the associated EOF), and $GIA(\bm{x},t-t_0)$ represents the GIA term (implemented through a correction from a single, selected GIA model).  The solution, including the amplitudes of the leading EOFs, models the change in RSL from one time step to the next (e.g., monthly averages for \citealp{Church2004Estim,Church2011}). 

An advantage of EOFs is that they learn about correlations from the observations, which allows for complex patterns with teleconnections. The assumption that dominant spatial patterns are constant over time, across frequencies of variability, and over the changing selection of tide gauges may lead to biases, however, because of the sensitivity of EOFs to the choice of spatial domain and time period. Additionally, features of physical modes can be mixed between EOFs, and there is no guarantee that an EOF pattern has physical meaning; instead, the patterns can represent noise (see \citealp{Calafat2014EOF} for a detailed critique).

\subsubsection{Incorporating physics-based models}
More physical knowledge can be incorporated at the process level by building physics-based models into the covariance structure or by using a probabilistic ensemble approach.  For example, rather than optimizing hyperparameters of a covariance function based on the data, \cite{Kopp2009probabilistic} used physical models of glacial-isostatic adjustment (GIA) to help define the prior covariance structure of a spatio-temporal GP for an analysis of GMSL and RSL change during the Last Interglacial.  This approach still assumed a GP prior; more complex priors can be represented through more direct use of physics-based models or emulators thereof.

Although the analysis methods (Section \ref{S:Analysis}) used in the implementations differ, the process models of \cite{Hay2015probabilistic} and \cite{Dangendorf2017GMSL}, used to analyze the instrumental record, are similar variants of:

\begin{equation}f(\bm{x},t)=g(t)+\Sigma_j{FP_j(\bm{x})M_j(t)}+DSL(\bm{x},t)+GIA(\bm{x})(t-t_0)+NL(\bm{x},t)+w(\bm{x},t).
\label{eq:KSProc}
\end{equation}

\noindent Here, the spatio-temporal RSL field is split into several component fields. A globally uniform term, $g(t)$, includes global thermal expansion and unmodeled sources of change. (It is not, however, representative of GMSL, as several of the other terms have non-zero global means). $FP_j$ and $M_j$ are the static-equilibrium fingerprint and melt, respectively, for each ice sheet/glacier source regions, indexed by $j$. $DSL(\bm{x},t)$ is dynamic sea-level change, estimated using information from atmospheric/ocean global climate models.  $GIA(\bm{x})$ is the local contribution to RSL from GIA, estimated using information from GIA process models, and $w(\bm{x},t)$ is process noise. 

An advantage to incorporating knowledge of processes through physical models is that they add potential information in the open ocean, far from tide gauge sites, whereas purely statistical models lose power away from the data. They also allow for teleconnections, rather than assuming informativeness always decreases with distance. A disadvantage is that they can be more complex to implement and may be overly rigid and rely on a small number of interpretations of physical processes.  Current implementations also use discrete inputs and outputs, without comprehensively accounting for uncertainties in the parameters that determine the process or using a continuous parameter space.

Emulation of complex physical models, including 1-D and 3-D GIA models and ice-sheet models, with statistical models (e.g., Gaussian process emulators; \citealp{Kennedy2006GPEmul,Rougier2008GPEmul}) or simplified physical models (e.g., \citealp{Urban2010SimpleMods}) can provide a faster, more flexible way of explicitly embedding this knowledge in a hierarchical framework. Statistical emulation reduces the processing time of these physical models, which are computationally intensive; it produces continuous output, in contrast to the discrete sea-level curves that are output for each set of discrete input parameters; and it enables probabilistic conclusions about the input parameters driving the physical models. 

\section{Analysis choices}
\label{S:Analysis}
Analysis methods used in the sea-level modeling literature include least-squares analysis (e.g., \citealp{Church2004Estim,Shennan2002GB,Engelhart2009Geol}), \textit{ad hoc} approaches such as `virtual station' averaging (detailed in section \ref{SS:ad_hoc}; e.g., \citealp{Jevrejeva2006,Dangendorf2017GMSL}), empirical Bayesian analysis (e.g., \citealp{Kopp2009probabilistic,Khan2017Carib,Meltzner2017Bel}), fully Bayesian analysis (e.g., \citealp{Parnell2005thesis,Cahill2016bayesian}), Kalman smoother (KS) algorithms (e.g., \citealp{Hay2015probabilistic}), and direct and approximate methods for incorporating temporal uncertainty in proxy data into statistical models.  Simple process models can be implemented with almost any analysis choice, while more complex models may require non-linear least squares or a Bayesian approach.

\subsection{Least squares}\label{SS:LS}
Least-squares analysis optimizes a model by minimizing the sum of squared deviations between the observed RSL and a RSL-process model function \citep{Legendre1805LS}. It can be used with functions as simple as a line (i.e., simple linear regression, Section \ref{SS:LR}) or as complex as in EOF regression (Section \ref{SS:emp_orth}).  Least-squares analysis is included with most statistical software (e.g., R, MATLAB, SAS) and is easy to implement with many modeling choices.  However, ordinary least-squares (OLS; \citealp{Aitken1934GLS}) analysis does not include implementation of a data level, and therefore typically excludes explicit measurement and inferential uncertainties. It also assumes errors are independent and identically distributed.

Slightly more advanced solutions than OLS include weighted least squares (WLS) and generalized least squares (GLS).  WLS addresses the problem of \textit{heteroscedastic} (unequal) variances, and GLS additionally addresses the problem of autocorrelation among variances, both of which are common characteristics of data used in sea-level analyses.  GLS estimators can be more efficient than OLS estimators \citep{Goldberger1962GLS}.  OLS, WLS, and GLS all require parametric linear models (though note that a linear model need not be a linear function of time).  Total least squares \citep{Golub1973TLS,Golub1999TLS} is a generalization of the least-squares approximation method and incorporates uncertainty in both the independent and dependent variable, and non-linear least squares uses optimization algorithms to maximize the fit of more complex models.

\subsection{Ad hoc approaches: regional averaging, virtual stations, pre-processing}
\label{SS:ad_hoc}
\begin{figure}[ht]
\centering\includegraphics[width=0.85\linewidth]{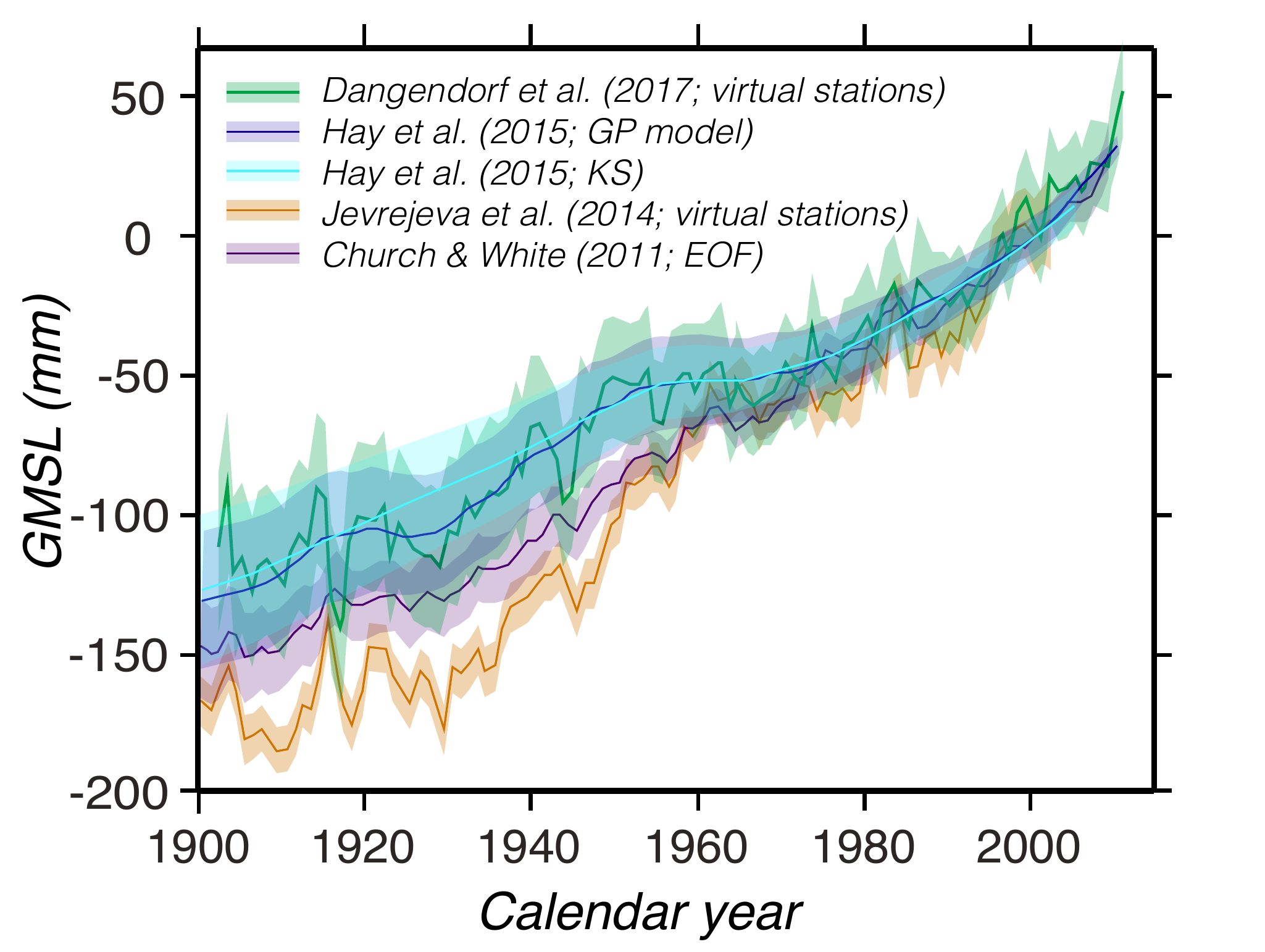}%
\caption{Comparison of GMSL curves based on different subsets of tide gauges, process model choices, and analysis methods, including KS, GP, virtual stations, and EOF regression \citep{jevrejeva2008recentaccel,Church2013ERL,Hay2015probabilistic}.}
\label{fig:GMSL}
\end{figure}

We define `ad hoc' approaches as analysis methods constructed without an underlying statistical theory.  Modern estimates of GMSL change apply various versions of these ad hoc approaches, including regional averaging, `virtual stations,' and pre-processing to different subsets of tide gauges.  The results of these techniques exhibit various GMSL curves (Figure \ref{fig:GMSL}).

Regional averaging effectively removes the contributions of some processes, such as those included in the regional and local terms of purely statistical models.  Definitions of the number of regions and how the averaging is implemented vary by study.  \cite{Jevrejeva2006,Jevrejeva2009GRL26417,Jevrejeva2016PNAS} attempted to address the spatial heterogeneity of tide gauges separated by geographic regions through a `virtual station' approach, which iteratively averages rates between stations to estimate a regional average and then averages across all regions to find a global average (Figure \ref{fig:GMSL}). \cite{Dangendorf2017GMSL} adopted the general idea of Jevrejeva's `virtual station' technique and weighted each regional estimate by its approximate area in relation to the entire ocean (Figure \ref{fig:GMSL}). 

Regardless of the model, many analyses `correct' for physical processes prior to analysis (e.g., \citealp{Cahill2015modeling,Tamisiea2011GeogVar,Church2011}) by removing sites that do not meet desired criteria and by subtracting out signals from physics-based process models prior to analysis (i.e., pre-processing).  For example, within regional averaging implementations, \cite{Douglas1991GSLR,Douglas1997GSL} and \cite{Holgate2007GRL} corrected for the effects of GIA using single GIA models and screened out tide-gauge stations deemed to include a sizable tectonic contribution. \cite{Dangendorf2017GMSL} corrected each tide gauge, prior to analysis, according to the static-equilibrium fingerprints of assumed melt components, GIA, and vertical land motion, which were each estimated by physical process models. For more details on the `virtual station' approach, see Section \ref{SI:RA}.

\subsection{Empirical Bayesian analysis}\label{SS:EHM}
Empirical Bayesian analysis, employing \textit{Empirical Hierarchical Models} (EHMs; see \citealp{Cressie2015statistics,Gelman2013Bayesian} for background), uses point estimates of the parameters based on the RSL data (e.g., \citealp{Kopp2016PNAS,Hay2015probabilistic}).  Maximum likelihood estimates (MLEs, $\hat{\theta}$) are optimal point estimates found by maximizing the likelihood $p(y|\theta)$ of the model, given fixed data. An EHM yields a posterior distribution of RSL, conditional on the data and the optimal parameters $p(f|y,\hat{\theta}_s,\hat{\theta}_d)$.  Although explicit bounds are usually set on hyperparameters for MLEs, there is no explicit prior distribution on the parameters.  Instead, the parameter level describes the optimization or estimation of the data and process parameters, $\theta_d$ and $\theta_s$, respectively.

\begin{equation}
\underbrace{p(f|y,\hat{\theta}_s,\hat{\theta}_d)}_{\text{posterior}}\propto \underbrace{p(y|f,\hat{\theta}_d)}_{\text{likelihood}}\cdot \underbrace{p(f|\hat{\theta}_s)}_{\text{prior}}.
\end{equation}
Almost all published implementations of RSL process models with spatio-temporal GP priors applied to RSL proxy data use empirical Bayesian analysis (e.g., \citealp{Kopp2015NC,Khan2017Carib}). For instrumental data, \cite{Hay2015probabilistic,Hay2017FP} demonstrated an EHM with GP priors alongside the KS approach (section \ref{SS:KST}) to estimate GMSL, the spatio-temporal RSL field, and the components contributing to RSL globally at decadal intervals from tide gauge records.  \cite{Meltzner2017Bel} implemented an empirical GP model using coral microatoll proxy data from the mid-Holocene in Southeast Asia to estimate rates of RSL change by incorporating a periodic term to capture the 18.6-year tidal cycle.   

\begin{figure}[ht]
\centering\includegraphics[width=0.85\linewidth]{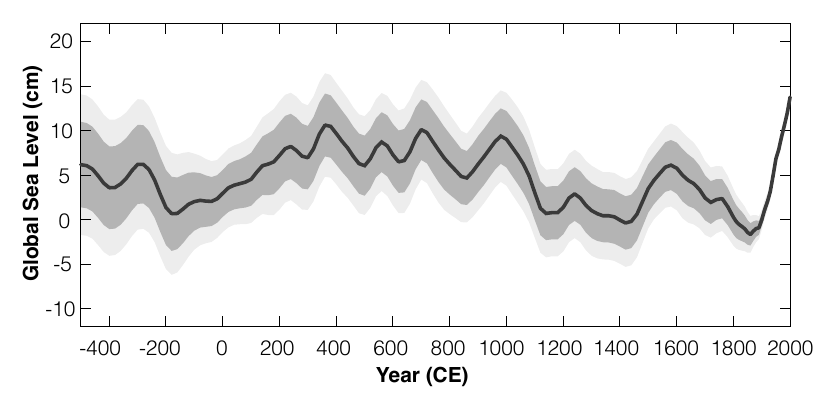}%
\caption{GMSL estimate with 67\% and 90\% credible intervals over the last $\sim$2500 years from \cite{Kopp2016PNAS} using a model with GP priors, applied to RSL proxy data and instrumental data in an empirical Bayesian analysis.}
\label{fig:GMSL_CE}
\end{figure}

GMSL reconstructions fusing proxy and instrumental data are possible using empirical Bayesian analysis, although they have rarely been implemented. \cite{Kopp2016PNAS} provide the only example of using both instrumental and proxy data to construct an empirical GMSL reconstruction over the past 2500 years using spatio-temporal modeling with empirical Bayesian analysis (Figure \ref{fig:GMSL_CE}).  

EHMs generally require fewer computational resources than fully Bayesian techniques; however, like fully Bayesian approaches, empirical GP implementations require computation of the inverse of a full covariance matrix (over all times and space), the computational demands of which are more strenuous than a state-space model (a model that is defined by a system of first-order difference equations of state variables), which estimates a covariance matrix at each time step. For this reason, EHM analyses (and BHM analyses) do not scale to large datasets as easily as other approaches.

\subsection{Fully Bayesian analysis}
\label{SS:BHM}
Another analysis choice, fully Bayesian analysis, gives rise to Bayesian Hierarchical Models (BHMs, see \citealp{Cressie2015statistics,Gelman2013Bayesian} for background). A fully Bayesian analysis requires that all model parameters have prior probability distributions, allowing parameters to take on a range of probable values.  These prior distributions may incorporate informative prior knowledge or may be uninformative, vague priors.  Priors are typically sampled using \textit{Markov Chain Monte Carlo} techniques (MCMC: algorithms used to approximate random samples from complex probability distributions, e.g., \citealp{MCMC2011}); however, for a limited set of likelihood and \textit{conjugate prior} distributions, combined with relatively simple model structures and known hyperparameters, they can be solved analytically.

The output of a BHM is the posterior distribution, $p(f,\theta_s,\theta_d|y)$, of the sea-level process, $f$ (e.g., the probability distribution of RSL across time and space), and the parameters, $\theta_s$ and $\theta_d$, given the observed data, $y$.  This posterior is proportional to the product of the likelihood of the model, $p(y|f,\theta_d)$, the prior distribution of the model, $p(f|\theta_s)$, and the prior of the parameters, $p(\theta_d,\theta_s)$, where $\theta_d$ and $\theta_s$ are the data and sea-level process hyperparameters, respectively:

\begin{equation}
\underbrace{p(f,\theta_s,\theta_d|y)}_{\text{posterior}}\propto \underbrace{p(y|f,\theta_d)}_{\text{likelihood}}\cdot \underbrace{p(f|\theta_s)\cdot p(\theta_d,\theta_s)}_{\text{prior}}.
\end{equation}

As with empirical Bayesian analysis, fully Bayesian analysis can be implemented with virtually any process model (e.g., \citealp{Parnell2015SVM,Cahill2015CP,Cahill2016bayesian,Piecuch2017JRG}). In general, it is more computationally demanding than an empirical analysis but provides more thorough estimates of relative uncertainties (e.g., \citealp{Piecuch2017JRG}).

\subsection{Kalman smoother techniques}
\label{SS:KST} \label{SS:KS}
The Kalman smoother can combine process-based models of the drivers of sea-level change with spatially and temporally sparse observations to estimate a temporal or spatio-temporal model. Implementation of the KS is based on the Kalman filter \citep{Kalman1960KS}, a data assimilation technique that iteratively performs a least-squares analysis whenever observations are available, but in the absence of observations relies on model dynamics to compute the best estimate of the state vector.  The Kalman filter method assumes that the state at time $k$ evolves linearly from the state at $k-1$. The approach is similar to a Bayesian updating process, occurring one time step at a time, and is equivalent to a linear Gaussian state-space model. (For examples of non-linear, non-Gaussian state-space models, see \citealp{Parnell2015SVM,Cahill2016bayesian}).  The KS extends the Kalman filter so that estimates at any given point in time are informed by observations in its future as well as in its past.  For example, \cite{Hay2013GSL,Hay2015probabilistic} used the Kalman smoother to implement a model similar to that in equation \ref{eq:KSProc} and thus model GMSL, the field of RSL change, and different driving processes (see Section \ref{Sup:KS} for more details on this implementation).

The KS approach is flexible in terms of process models that can be represented. Because it is recursive, it is computationally faster than approaches (such as EHMs or BHMs) that require estimating all spatio-temporal points simultaneously; the KS scales linearly with the number of data points \citep{Grewal2001KS}. It is therefore especially valuable for estimating GMSL and RSL over the instrumental period, as it enables analysis of data at a higher temporal resolution than non-recursive analysis methods with comparable modeling choices. However, the KS approach fails for low data density (which can be shown analytically; see \citealp{Hay2017FP,Gelb1974Interp}), does not readily handle temporal uncertainty \citep{Kalman1960KS,Visser1988KF}, and therefore has not yet been implemented in the literature using proxy data to estimate trends over longer timescales. Alternative methods have been used to incorporate temporal uncertainties within models with GP priors, including errors-in-variable (EIV) and \textit{noisy-input GP (NIGP)} methods (Section \ref{SS:Temp_Imp}); however, the KS has not been applied to data with temporal uncertainties.

\subsection{Incorporation of temporal uncertainty} \label{SS:Temp_Imp} \label{SS:EIV}

The temporal uncertainty of RSL proxy data has been incorporated into models in various manners. An EIV framework, which has been implemented in temporal IGP (e.g., \citealp{Kemp2013NJ}, \citealp{Gehrels2013Modern}, \citealp{Brain2015NC}, \citealp{Cahill2015modeling}) and change-point (e.g., \citealp{Cahill2015CP}) models and in a spatio-temporal GP model \citep{Kopp2009probabilistic}, incorporates temporal uncertainty directly through MCMC sampling of the distributions.  Because of its use of MCMC sampling, the EIV framework is generally employed together with a fully Bayesian analysis approach.

An alternative approach with less computational demand is to approximate and recast temporal noise as RSL uncertainty. The noisy-input Gaussian Process (NIGP) method of \citealp{McHutchon2011Noisy}) has been implemented to do this in temporal and spatio-temporal empirical models with GP priors (e.g., \citealp{Miller2013SLNIGP,Kopp2015NC,Khan2017Carib}.  The NIGP uses the first-order Taylor-series approximation -- a linear expansion about each input point -- to translate errors in the independent variable, time, into equivalent errors in the dependent variable, RSL, such that temporal uncertainty is recast as sea-level uncertainty proportional to the squared gradient of the GP posterior mean. 

\section{Illustrative Analyses}
\label{S:Implement}

There are a number of modeling and analysis choices that can be used to evaluate a given scientific question.  To illustrate the advantages and disadvantages of specific implementations in RSL modeling, we apply several models to common datasets. We focus on pairs of modeling and analysis choices that commonly occur together in the literature and organize them from simple to more complex.  First, we demonstrate the differences between several time-series models -- temporally linear models with least-squares analysis (e.g., \citealp{Shennan2002GB,Engelhart2009Geol}), change-point and EIV-IGP models with fully Bayesian analyses (e.g., \citealp{Brain2015NC,Cahill2015CP,Cahill2016IGP}), and temporal models with GP priors with empirical Bayesian analysis (\textit{ET-GP}) -- to estimate RSL change from continuous cores over the Common Era. The quasi-linearity of RSL over this period warrants an evaluation of linear and change-point models. Next, we demonstrate a spatio-temporal GP model with empirical Bayesian analysis (\textit{EST-GP}; e.g., \citealp{Khan2017Carib,Kopp2016PNAS}), which characterizes spatial and temporal variability in RSL change over the Holocene using proxy data. This is the only technique currently used in the RSL literature that accommodates both temporal uncertainties and spatial correlations. Therefore, to illustrate the spatio-temporal approach, we compare the results of the EST-GP model to a site-by-site ET-GP. Last, we analyze tide-gauge data with a physically-informed KS model (e.g., \citealp{Hay2013GSL}) and an EST-GP model (e.g., \citealp{Hay2017FP}), as they are capable of estimating both GMSL and the spatio-temporal fields of RSL and its rates of change with uncertainties.  The type of data, time period of interest, and relevant scientific question determines which techniques are appropriate. Table \ref{t:mods} outlines the implementations applied in Sections \ref{SS:CE}, \ref{SS:CompHol}, and \ref{SS:Comp1900}), and the detailed descriptions of each model can be found in Section \ref{SI:Imp}.

\begin{table}
  \tiny
  \caption{Illustrative analyses table}
 \label{t:mods} 
 \begin{tabular}[l]{m{3.7cm} m{2.4cm} m{6.0cm} m{1.5cm} m{1.3cm}}
\textbf{Implementation} &\textbf{Analysis Approach} &\textbf{Assumptions} &\textbf{Illustrative Dataset} &\textbf{Run Time} \\ [0.5ex]
  \midrule
Time series analyses		&&&& \\								
  \midrule
{Temporally linear regression} & {	general least-squares	} & {	Linear signal, Gaussian uncertainties, does not distinguish measurement uncertainty and process nonlinearity	} & {	Continuous core	} & {2 seconds	} \\
{Change-point regression} & {	fully Bayesian	} & {	Segment-wise linear, Gaussian uncertainties, does not distinguish measurement uncertainty and process nonlinearity	} & {	Continuous core	} & {	24 minutes	} \\
 {Empirical temporal Gaussian process model (ET-GP)}  & {	empirical Bayes	} & {	Gaussian uncertainties, Stationary covariance of sea level correctly characterizes temporal variability, Once-differentiable sea-level signal	} & {	Continuous core	} & {	60 seconds	} \\

{ Bayesian error-in-variable integrated Gaussian process model (EIV-IGP)	} & {	fully Bayesian	} & {	Gaussian uncertainties, Stationary covariance of sea-level rate (can lead to non-stationary sea level covariance) correctly characterizes temporal variability, Twice-differentiable sea-level signal	} & {	Continuous core	} & {	6.1 hours	} \\
  \midrule
Spatio-temporal analyses				&&&& \\						
  \midrule
 {	Empirical spatio-temporal Gaussian process model (EST-GP)	} & {	empirical Bayes	} & {	Spatio-temporal covariance correctly characterizes variability	} & {Holocene
Tide gauge} & {1.2 hours
26 minutes} \\
{	Linear state-space model (KS)	} & {	Kalman smoother	} & {	Based on the physics-based models used; linear in prediction from state to state	} & {	Tide gauge	} & {	4 hours	} \\
\midrule
\end{tabular}
Details pf the analyses performed in Section \ref{S:Implement} and the run times of each implementation on a standard laptop computer.
\end{table}

\subsection{Estimating rates of RSL change from continuous cores (Common Era)}
\label{SS:CE}\label{SS:CompCE}

Attempting to answer scientific questions about the timing of RSL accelerations in relation to climatic drivers requires RSL proxy data because instrumental records are frequently too short. The near-continuous records from single cores of salt-marsh sediment (continuous cores) are well-suited to capturing the onset of modern rise because they provide a longer record than instrumental data, and they possess sufficient vertical and temporal resolution to provide a meaningful estimate of the timing of accelerations in sea-level rise. The data used in the following analyses include previously published data from continuous cores collected at two sites in New Jersey (Leeds Point and Cape May Courthouse; \citealp{Kemp2013NJ}, 135 data points) and one site in North Carolina (Sand Point; \citealp{Kemp2011PNAS}, 109 data points), where the New Jersey sites are assumed to be independent of the North Carolina site. We applied the linear, change-point, ET-GP, and EIV-IGP models, described in Section \ref{SI:TS_Mods}, to the data (Section \ref{SI:Dat}).

\begin{sidewaysfigure}
\centering\includegraphics[width=1.0\linewidth]{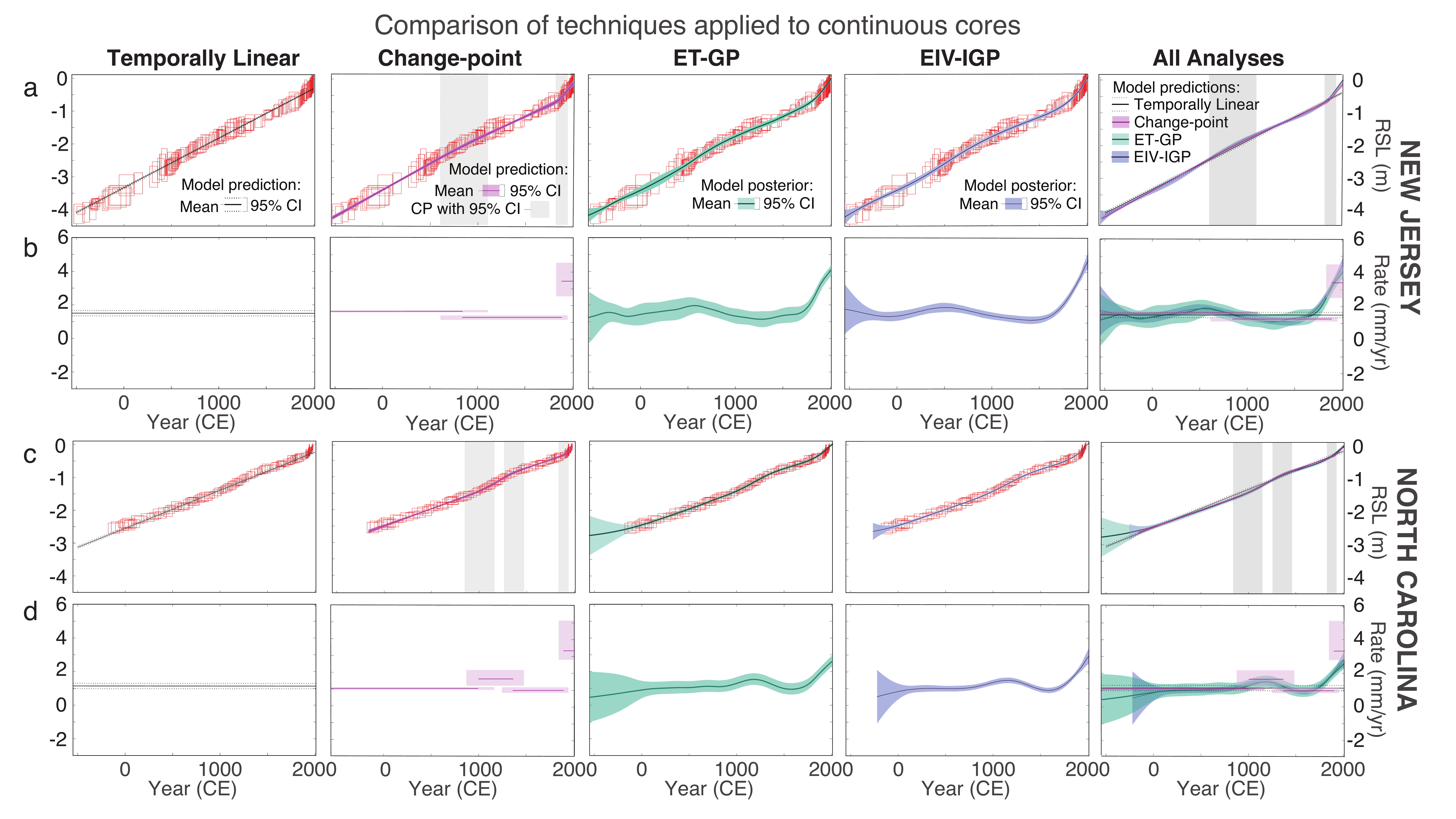}
\caption{Common Era comparison of linear model results (column 1), change-point analysis (column 2), temporal empirical GP model results (column 3), and Bayesian EIV-IGP model results (column 4), where input data are continuous cores, and there is no spatial correlation, in New Jersey (a,b), and North Carolina (c,d).  Output includes estimates of RSL (a,c) and rates of RSL change (b,d), which are each shown with mean and 95\% confidence or credible intervals, where each model covers more than 95\% of the data, except for the linear regression model.}
\label{fig:comp}
\end{sidewaysfigure}

We show estimated RSL and rates of RSL change for each model in Figure \ref{fig:comp}.  The assumptions in these models determine their form such that the incorporation of temporal and vertical uncertainty within the change-point, ET-GP, and EIV-IGP models more appropriately characterize the uncertainties of the data. Conversely, the rigidity of the temporally linear model does not accommodate the underlying process(es) influencing the data.  The ET-GP and Bayesian EIV-IGP models yield similar mean estimates, although the Bayesian EIV-IGP makes somewhat more precise predictions with smaller uncertainties (Figure \ref{fig:comp}) in this particular application, which may be due to the choice of covariance functions or their hyperparameters (see Section \ref{sup:cov} for discussion of the details of covariance functions and their parameters).

\begin{figure}[ht]
\centering\includegraphics[width=1.0\linewidth]{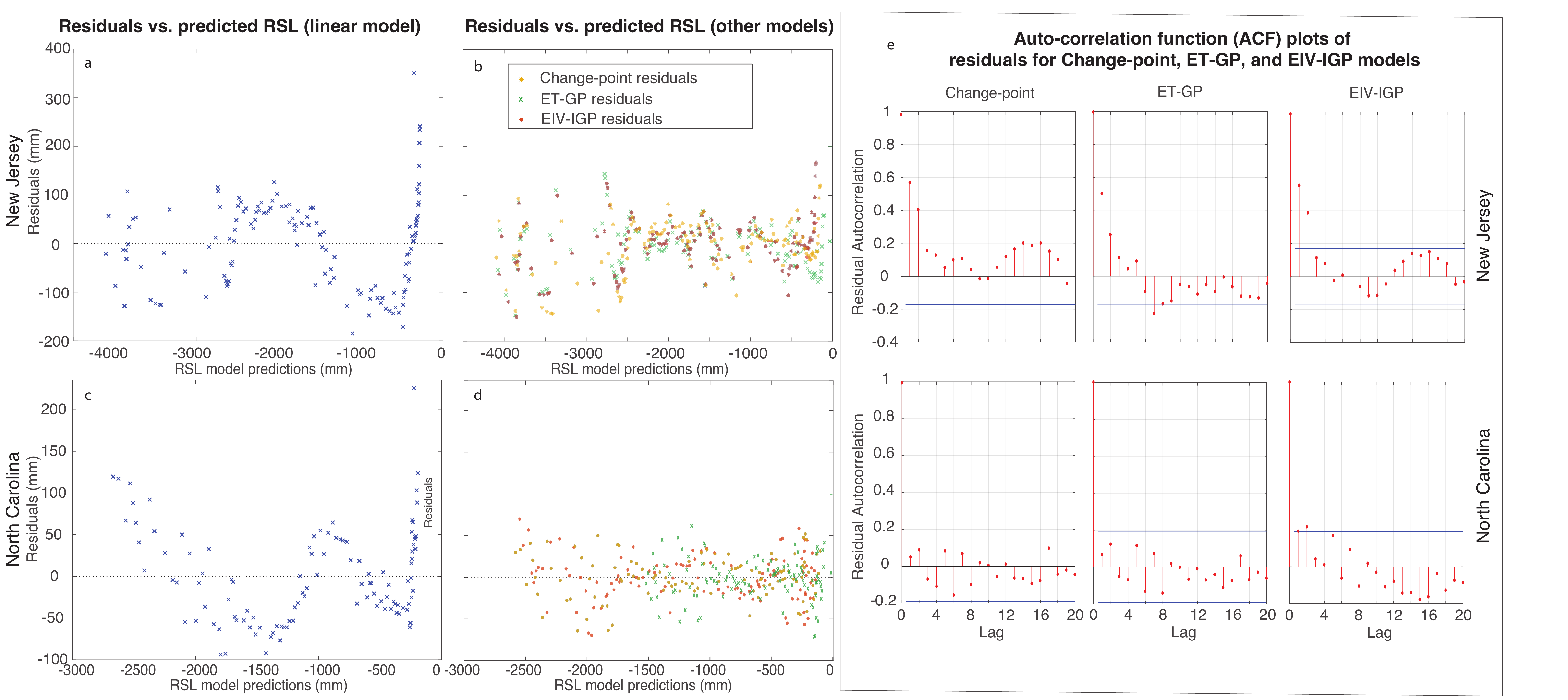}%
\caption{Linear model (a,c) and all models (b,d) residuals plotted against the predicted RSL values for each site (New Jersey: a,b, North Carolina: c,d). (e) Autocorrelation function plot of residuals for each model, where the EIV-IGP North Carolina model and all three New Jersey models violate the independence assumption, because they show significant autocorrelations (more than 5\% of which are above or below the blue 95\% significance line).}
\label{fig:Resid}
\end{figure}

Each of the models implemented assume that the data error terms in the models are independent, while the ET-GP model assumes those errors are normally distributed, and the temporally linear model pools all data error terms and assumes errors are independent and identically distributed (iid).  In order to test these assumptions (see Sections \ref{S:TimeSeries} and \ref{S:Analysis} for a more thorough description of model assumptions), we present a plot of the residuals as a function of the predicted RSL values and an autocorrelation function (ACF) plot, which shows the autocorrelation in residuals as a function of lag, the number of time steps between predictions (Figure \ref{fig:Resid}).  If all assumptions are met, the errors should be random, meaning there is no pattern or correlation in the residuals.  The residuals of the linear regression model display a non-random, temporally-dependent pattern, indicating that the model does not fit this dataset well and the temporally linear model is inconsistent with the temporal evolution in the data. The non-linear models show less structure in the residuals, as well as smaller residuals than the linear model. In addition, one risk in change-point analysis as well as models with GP priors is that they may violate the assumption of independent errors.  An ACF plot can determine whether this assumption is met or violated.  For example, the ACF plot in Figure \ref{fig:Resid} demonstrates that the analyses of the North Carolina dataset do not violate the independence assumption.  Conversely, the change-point, ET-GP, and EIV-IGP analyses of the New Jersey dataset do, shown by the significant (at 5\% significance level; blue lines in Figure \ref{fig:Resid}) autocorrelation in residuals. This signifies that the residuals may contain additional information that is not included in the model structure.

Some of the differences between the ET-GP and EIV-IGP models are associated with the conventional choices of covariance functions used.  The squared-exponential covariance function, used in the EIV-IGP, is slightly smoother in this implementation than the Mat\'ern covariance function (see Section \ref{sup:cov}) in the ET-GP (Figure \ref{fig:comp}). However, the choice of time length over which the rate of RSL change is averaged in the ET-GP model affects the smoothness of (and uncertainty in) the rate curve; A linear transformation on the predicted RSLs is performed to calculate the rate curve and its uncertainty.  The ET-GP method, as employed, enables more complexity for various processes through multiple separate covariance functions, summed together to create the composite RSL process, but either method could incorporate various covariance functions.

In addition to making different assumptions, these four implementations produce distinct results about the probability and timing of accelerations in RSL.  A temporally linear model will never predict an acceleration in RSL because of its inherent assumption of a constant rate of RSL change, whereas a change-point model is designed to detect slight changes in rate, but assumes instantaneous acceleration.  The ET-GPR and EIV-IGP, conversely, produce continuous posterior distributions on rates of RSL change over time.  Any inflection points in the rate curves (Figure \ref{fig:comp}b,d) can be interpreted as changes in trend, but their significance must be evaluated. For example, the ET-GP model estimates a significant (at 5\% significance level) difference in the rate of RSL change in New Jersey between 1270 and 1795. Inflections can also be observed in the rate curve for the New Jersey record around -170 and 570 CE, indicating changes in rate, but the differences are not significant.  Alternatively, taking the derivative of the rate curve, for either the ET-GP or EIV-IGP model, would produce a probabilistic estimate of acceleration (or deceleration) over time. Hence, the non-parametric nature of the ET-GP and EIV-IGP leads to more flexible inference about the evolution of RSL. 



\subsection{Characterizing spatial and temporal variability in RSL change from proxy data (Holocene to present)}

Attempting to answer scientific questions about the regional-scale patterns of RSL associated with different driving physical processes requires combining information from various sources and locations in a spatio-temporal model.  We apply the EST-GP to proxy data to illustrate the only model in the literature that accommodates both temporal uncertainties and spatial correlations. 

We compiled data from previously published studies \citep{Engelhart2012QSR,Kemp2013NJ,Kemp2014MGLate,Kemp2015RSLConn,Kemp2017RSLNYC,Kemp2017QSR,Khan2017Carib} along the Atlantic coast of the United States and the circum-Caribbean (latitudes 24.95$-$44.68 $\degree$N, longitudes 67.38$-$81.73$ \degree$W) from 11 ka to present (Supplemental \ref{SI:Dat}).  We employ 450 SLIPs spanning from 8 ka to present from \cite{Engelhart2012QSR}, 66 SLIPs from 11 ka to present from \cite{Khan2017Carib}, and 498 continuous core data points and 28 SLIPs from 3 ka to present from \cite{Kemp2013NJ,Kemp2014MGLate,Kemp2015RSLConn,Kemp2017RSLNYC,Kemp2017QSR}.  

We applied the EST-GP model to the whole spatio-temporal dataset, and also applied the ET-GP model on a site-by-site basis. The models implemented are described in Section \ref{SI:ST}. Other modeling and analysis choices reviewed within this manuscript (e.g., KS or EOF) have not been implemented in the literature with proxy data, and therefore are not included in this illustrative analysis.  

\begin{figure}[ht]
\centering\includegraphics[width=0.7\linewidth]{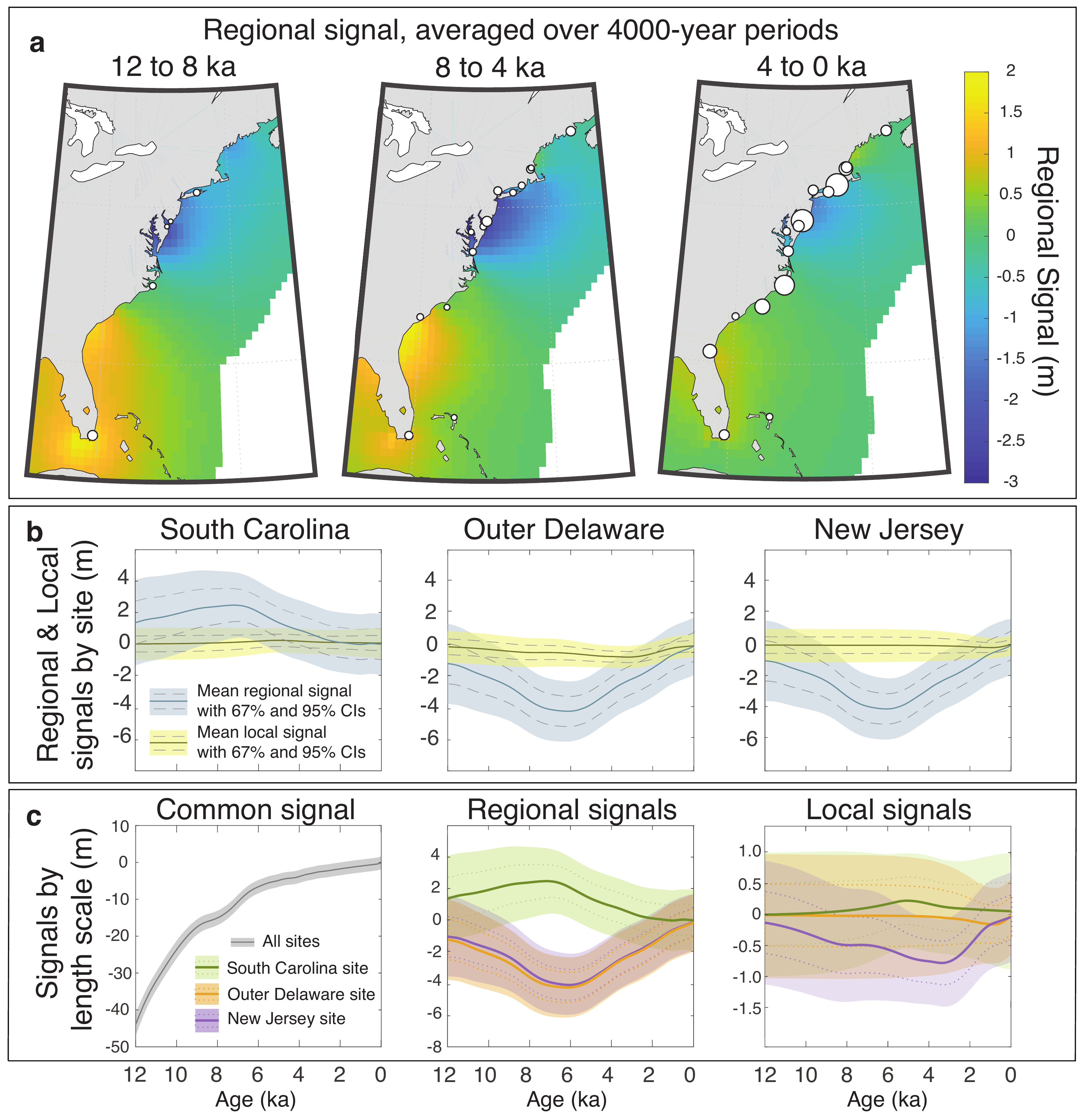}%
\caption{(a) Maps of the regional RSL signal, averaged over 4000-year periods, where the size of the white dots are proportional to the number of data points at each location, (b) the regional and local signals for three sites along the Atlantic Coast, (c) the common signal, which applies to the entire area of study as well as the regional and local signals at the same sites in b, plotted concurrently for comparison (note: these signals are not plotted on the same scale).}
\label{fig:Decomp}
\end{figure}

Whereas the ET-GP model only predicts RSL and rates of RSL change at sites with data (because the model is temporal only and runs independently for each site), the EST-GP model can make predictions at any point in space and time. Figure \ref{fig:Decomp}a shows the temporal evolution of RSL estimated by the EST-GP along the Atlantic coast of the U.S. and the geographic distribution of the data used in the model over time. Figure \ref{fig:Decomp}b and c show the temporal evolution of the rate of RSL change and uncertainties (standard deviation of the posterior RSL estimate), respectively. Figure \ref{fig:Decomp}d shows the locations of the sites used in the comparison between the spatio-temporal and temporal-only models (North Carolina; Figure \ref{fig:Decomp}e) and the cross-validation (Inner Delaware; Figure \ref{fig:Decomp}f), where site-specific data were omitted.  Despite the fact that all information for predicting RSL at this site comes from the correlation in RSL with other sites, where the data are only shown for comparison, the predictions fall very close to the data. Only 2 out of 28 data points fall outside of the 95\% credible interval model prediction of RSL. Generally, both ET-GP and EST-GP models are excessively conservative covering more than 95\% of the data that are used as input to the models within their 95\% credible intervals.  Figure \ref{fig:Decomp}g demonstrates the difference in posterior uncertainty estimates for two sites (shown on the map in panel c) with site-specific data (Outer Delaware) versus a site with no data near it (Merritt Island).

\begin{figure}[ht]
\centering\includegraphics[width=0.85\linewidth]{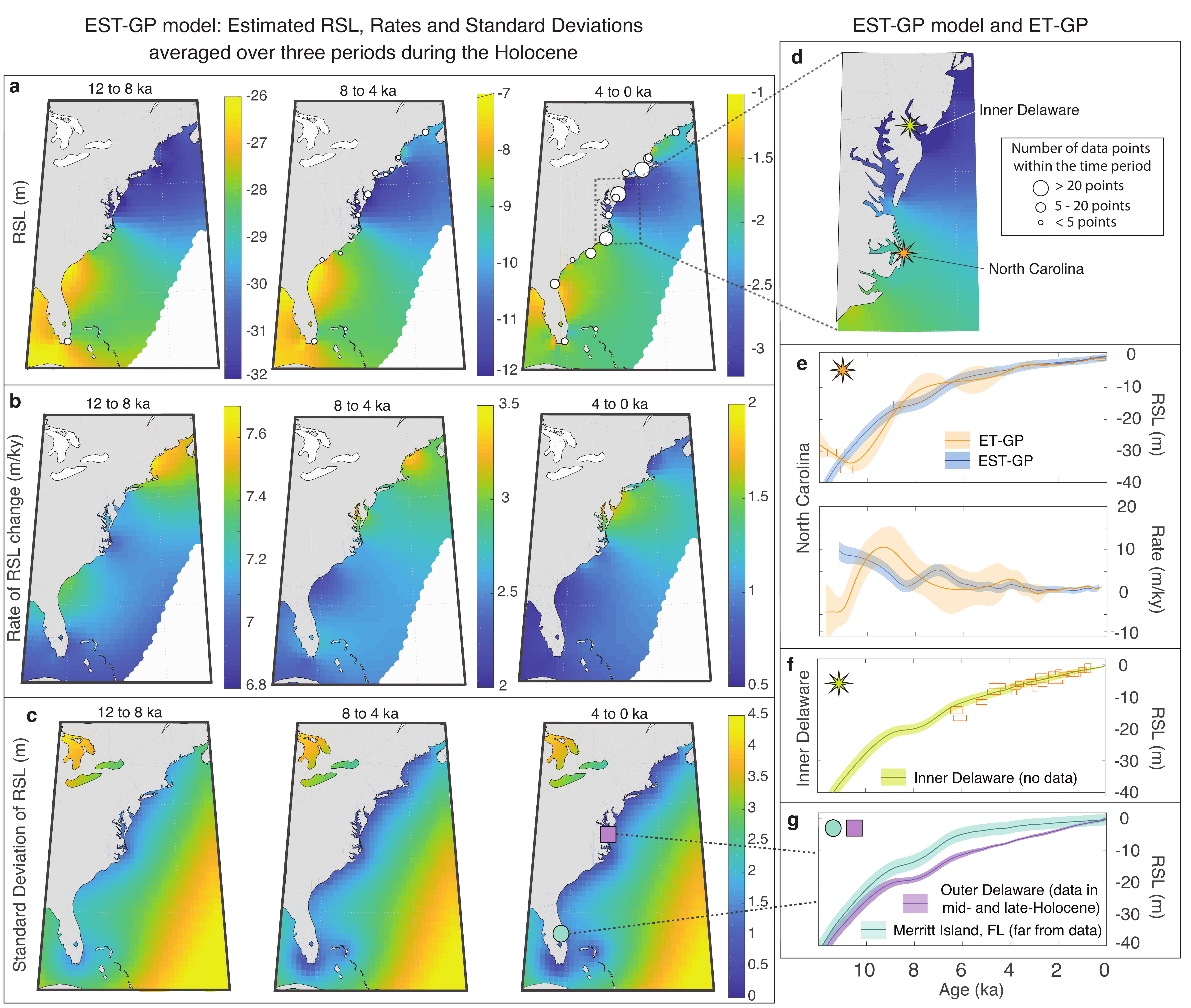}
\caption{Spatio-temporal Holocene model (EST-GP) results, including maps of mean estimated RSL (a), rate of RSL change (b), and standard deviation of RSL (c), averaged over 4000-year periods. (d) Stars on the map show the locations of two sites with estimated RSL curves (e,f). Comparison of predicted RSL and its rates of change are shown with 95\% credible intervals at North Carolina for the EST-GP (includes spatial correlations) and ET-GP models (e).  The EST-GP makes predictions at a site, Inner Delaware, where no data were used as input; data are shown for comparison purposes only (f). (g) Two models demonstrate the difference in uncertainties when the model is close to data (Outer Delaware, orange) versus far from data (Merritt Island, blue).}
\label{fig:Hol_IGPvGPR}
\end{figure}

\label{SS:CompHol}
One notable difference between the ET-GP and the EST-GP is the spatial correlation within the EST-GP model. The form of the RSL curve in North Carolina (Figure \ref{fig:Hol_IGPvGPR}e) is influenced by this correlation.  Whereas the ET-GP model produces higher RSL at 11 ka than 10 ka, the EST-GP uses information from the correlation with other sites to predict increasing RSL throughout the Holocene.  The EST-GP also maintains fairly constant uncertainties throughout the period of interest, whereas the ET-GP has less precision when data are sparse, due to the assumption of independence between sites.  However, at times and locations farther away from the data, the uncertainty increases in the EST-GP model, as well (Figure \ref{fig:Hol_IGPvGPR}c). Predicted uncertainty in RSL is greater at sites that are far from data (e.g., Merritt Island, FL; Figure \ref{fig:Hol_IGPvGPR}g) and in the early- to mid-Holocene (e.g., Figure \ref{fig:Hol_IGPvGPR}c), whereas uncertainties decrease by up to $\sim$80\% at times and locations with precise data (e.g., Outer Delaware; Figure \ref{fig:Hol_IGPvGPR}g).  The EIV-IGP model has not yet been applied to a spatial dataset and, along with the EST-GP, is not currently scalable to large datasets. Conducting the EST-GP with 5000 data points, on a standard laptop computer, for example, would lead to a computational time for a single model iteration of about 35 days; the fully Bayesian EIV-IGP analysis on the same dataset and same computational platform would – without improvements in computational efficiency – take about 1,300 years.

In these specific implementations, another notable difference is the process level model of the EST-GP, which has three distinct terms capturing common, regional, and local signals; the regional term for each site incorporates information from other sites within about 700 to 750 km, based on the optimized length scale parameter for that term, while the local term incorporates information from only 10 to 12 km distance.  These terms can be separated and analyzed (Figure \ref{fig:Decomp}), resulting in maps of the spatio-temporal signal for each term (Figure \ref{fig:Decomp}a) and plots of each term for specific sites. The common signal (which is uniform over the entire domain) absorbs a majority of the signal (Figure \ref{fig:Decomp}b), whereas the regional and local signals explain the variation between sites.  The higher RSL heights along the southern coast in the early- and mid-Holocene (12ka to 4ka)  are evident in the maps (Figure \ref{fig:Hol_IGPvGPR}) and in the regional curve for South Carolina (Figure \ref{fig:Decomp}b), which is shown to demonstrate distinctive patterns in regional signals in comparison to Outer Delaware and New Jersey. The Outer Delaware and New Jersey sites have lower regional signals, and the differences in these two sites is represented in the local signal (Figure \ref{fig:Decomp}b).  The regional term picks up differences among sites associated with GIA, a dominant regional process; however, the common signal absorbs a significant portion of the GIA signal, because of the similarity over the sites in the study area. Including a physical model at the process level may provide more insight into the relative contributions of other physical processes acting over different spatial scales.


\subsection{Estimating spatio-temporal RSL and GMSL from instrumental data (1900 to present)}

\begin{figure}[ht]
\centering\includegraphics[width=0.75\linewidth]{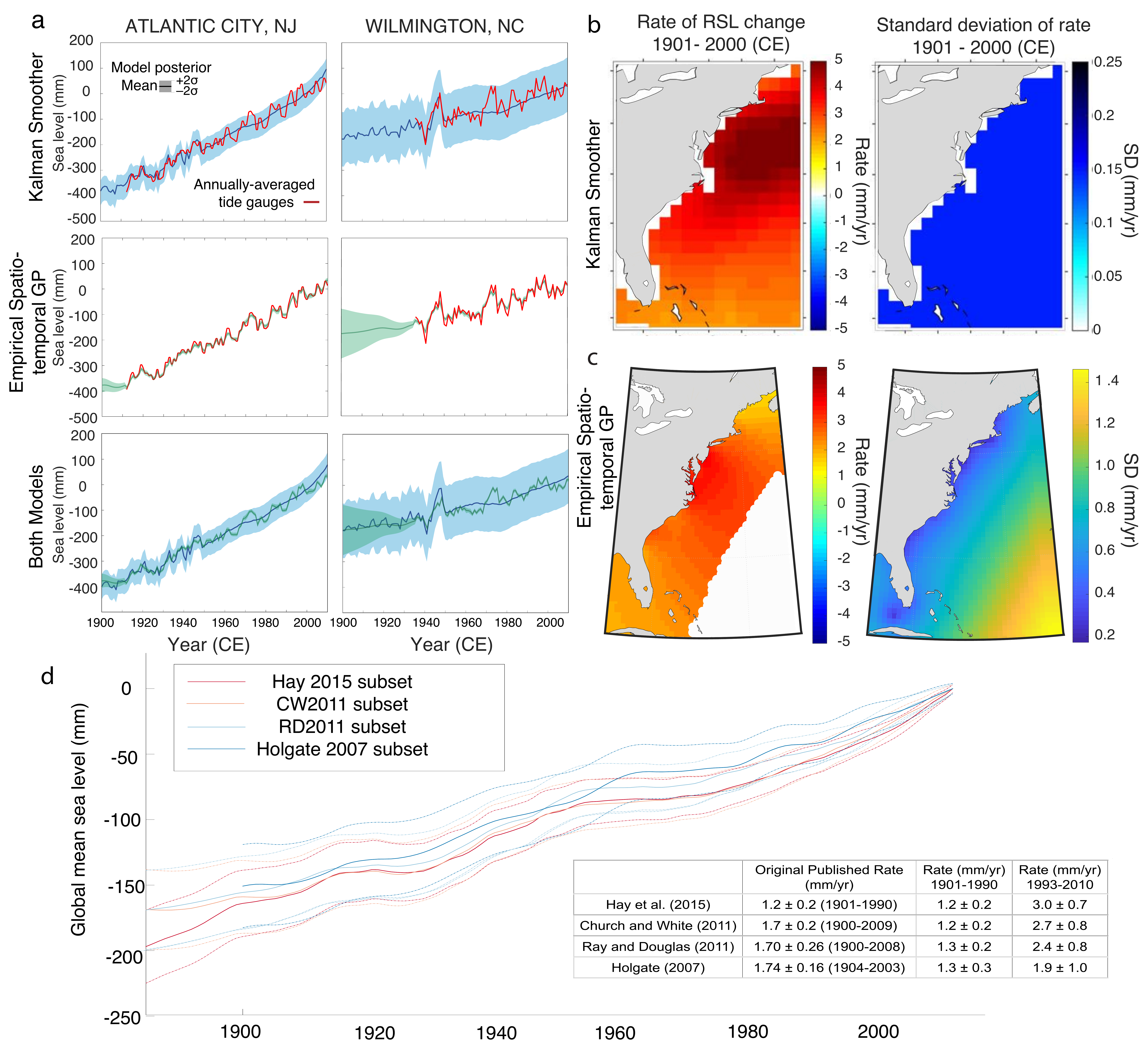}
\caption{Comparison of KS and EST-GP predictions at two sites (a) based on annually-averaged tide gauges, which are shown in red. Both models include a spatial component and produce maps of RSL rates of change, for the KS (b) and EST-GP (c). (d) GMSL time series obtained using the KS applied to subsets of tide gauges from previous studies, with $1\sigma$ uncertainties \citep{Hay2017FP}.}
\label{fig:KS_GPR}
\end{figure}

Attempting to answer scientific questions about GMSL change in the recent past requires instrumental records.  During the instrumental period, data include satellite altimetry measurements and tide gauges.  These data are inherently different from proxy data with negligible temporal uncertainty and smaller vertical uncertainties than proxies produce, and thus allow different methods.  Estimating GMSL through time and interpolating the spatio-temporal field of RSL change from instrumental records in the past are challenges well-suited to KS and GP model techniques.  Both techniques are implemented here using multiple tide-gauge records obtained from the Permanent Service for Mean Sea Level \citep{PSMSL2017,Holgate2013JCR}, with results shown at two sites: Atlantic City, New Jersey, (39.4$\degree$N, 74.4$\degree$W) and Wilmington, North Carolina (34.2$\degree$N, 78.0$\degree$W) (Figure \ref{fig:KS_GPR}). The models implemented are described in Section \ref{SI:instr}.
\label{SS:Comp1900}\label{SS:Comp1900AC} \label{SS:Comp1900MC}

Both techniques can compute posterior estimates of GMSL (Figure \ref{fig:KS_GPR}) as well as reconstruct the spatio-temporal sea-level field, conditioned on observed data, but their implementations are very different. The KS approach (Section \ref{SS:KST}; described in more detail in Supplemental \ref{Sup:KS}) steps through a forward filtering pass and a backward smoother pass for each time step, enabling computation of the covariance for a smaller subset of points and thus faster solution times ($\sim45$ seconds for a single KS run at the tide-gauge sites only, and $\sim4$ hours for the entire multi-model implementation globally). Conversely, the EST-GP conditions on all observations concurrently.  In \cite{Hay2015probabilistic}, both KS and GP implementations use output from physical process models.  However, in the current implementation, the EST-GP model has no physical-model input and is purely statistical, based solely on the data.  See \cite{Hay2015probabilistic} for a more complete treatment of these two analysis approaches.

We compare estimated RSL and uncertainty for the models at two sites in Figure \ref{fig:KS_GPR}a and the estimated rates of change in RSL and their uncertainties (standard deviations) in Figure \ref{fig:KS_GPR}b.  The spatial field computed by \cite{Hay2017FP} is less refined because of their modeling choice to compute the global field on a standard 5 degree grid.  A higher-resolution field can be computed with the KS; however, this will be accompanied by longer model run time.  Embedded in the KS spatial maps are dynamic sea level fields from several global climate models. When less data are available, for example earlier in time because the tide gauges are sparse, the KS predicts a much rougher sea-level time series for each location, despite the fact that there are tide gauges at these particular sites, whereas the EST-GP has larger uncertainties when there are no tide gauges as input at a specific site (Figure \ref{fig:KS_GPR}a).

Because of the differences in implementation of the KS and EST-GP, there are some drawbacks and advantages to each.  The inversion of the full covariance matrix (over all space and time points for data and predictions), which is required for the EST-GP makes the resolution of annual tide-gauge data difficult to handle when attempting to model these data on a global scale.  As a solution, lower-resolution (e.g., decadal) averages can be used as input \citep{Hay2015probabilistic}.  Conversely, the KS becomes unstable during the backward smoothing pass when persistent data gaps are present in the records.  The KS therefore requires a subset of tide gauges which ensures observation availability over time.

The KS model has also been tested on various subsets of tide gauges \citep{Church2011,Holgate2007GRL,Ray2011Exp}, which can slightly influence the results (Figure \ref{fig:KS_GPR}d, \citealp{Hay2017FP}). However, the results with a single analysis technique are more similar than when the data and analyses are both different.  When the research question relates to estimating GMSL, fully Bayesian methods may be too computationally intensive for the datasets; however, ad hoc (Section \ref{SS:ad_hoc}) choices may lead to different conclusions than the KS or GP techniques, especially when analyzing different sets of instrumental data.

\section{Discussion and conclusions}
\label{S:Conclusion}
Hierarchical statistical modeling frameworks are conceptual tools that provide a transparent approach for separating modeling choices at the data and process level from analysis choices. Appropriate modeling and analysis choices in sea-level research depend on the type of data and the scientific question(s) being addressed. We suggest that non-parametric approaches are generally best suited for analyzing RSL and rates of RSL change.

One goal in sea-level research is identifying changes in rates of RSL and GMSL change (e.g., \citealp{Church2006Accel,jevrejeva2008recentaccel,Kopp2013grl}).  We recommend temporal GP and IGP models, using fully Bayesian or empirical Bayesian analyses when the dataset includes instrumental or proxy data from a single site. These types of models enable probabilistic inference about accelerations and rates of change, and within a specific study, one can test hypotheses about the timing and magnitude of changes in rates.

Identifying the physical processes that explain patterns of spatial variability in RSL is a further objective of the sea-level community.  Spatio-temporal approaches are required to address this problem.  To date, spatio-temporal models of proxy data have generally used covariance functions that represent different scales of variability but do not tie these scales of variability to specific physical processes. Such models are useful for addressing this objective, but require post-inference comparison to physics-based models to interpret their results in terms of physical drivers. An alternative approach, yet to be demonstrated with proxy data, is to employ GP models that incorporate different processes through physics-based models (as in \citealp{Hay2015probabilistic}).  This approach could be powerful for certain processes (e.g., GIA or the elastic spatial fingerprints of ice-sheet mass changes) but in other cases (e.g., dynamic sea level change over multiple millennia) adequate physics-based models are lacking, so purely statistical process-level terms will remain necessary.  For constraining global sea-level change during the instrumental period, state-space analysis approaches, such as the KS, can manage large data sets more efficiently than empirical or Bayesian approaches that require simultaneously estimating all space-time points.

Improving estimates of GIA is a related goal to explaining spatial variability because it is the dominant driver of spatial variability in RSL change \citep{Peltier2015ICE6G} over mid-to-late Holocene timescales.  Traditionally, this is done through an iterative, manual process, where data from specific sites are compared to different versions of physical GIA models.  However, alternative approaches include using a suite or probabilistic ensemble of GIA models (e.g., \citealp{Hay2015probabilistic,Dusterhus2016LIG}) or using a single GIA model as a mean prior estimate and fitting the mismatch with a Gaussian process (e.g., \citealp{Kopp2016PNAS,VacchiprepRSLNEC}). Both approaches can permit further constraints on the role of spatially-variable GIA, while appropriately characterizing uncertainties, although the former approach requires that the truth be represented in the probabilistic ensemble or suite of models.

An important area of development for statistical sea-level models is more comprehensive and accurate use of data.  Most proxies do not conform to normal distributions, so techniques for incorporating non-Gaussian likelihoods, such as integrating transfer functions into spatio-temporal models, have the potential to make use of proxies that have been too approximately interpreted.  A general approach has been developed by \cite{Parnell2015SVM}, which readily applies to RSL.  Although they are usually carried out prior to process modeling, integrating transfer functions into full statistical models (e.g., \citealp{Cahill2016bayesian}) is a key goal for the next generation of palaeo-RSL models. Data compilation efforts (e.g., \citealp{Dusterhus2016ClimatePast}) aim to standardize and synthesize RSL data, which will enhance the comparability and accessibility of information to improve both physical models and statistical reconstructions. The accuracy and consistency of all statistical models will be enhanced when databases are standardized.  

A key methodological challenge is scaling spatio-temporal hierarchical modeling approaches for paleo-sea level data to large, yet still temporally noisy, datasets. Unlike Gaussian process models, whose computational complexity grows in proportion with the cube of the number of data points, the computational complexity of a Kalman smoother grows linearly. Adapting the Kalman smoother for temporally noisy data may provide an approach to overcoming these scaling problems and thus allow the simultaneous analysis of much larger proxy datasets. Recent work in the machine-learning literature exploring the translation between Gaussian process models and linear-Gaussian state space models (e.g., \citealp{Hartikainen2010GPKS}) may prove useful here. There are also several approximation and estimation techniques in the GP and machine-learning literature that have not yet been applied in a sea-level context, such as variational inference \citep{Blei2017VariationalInf}, which could speed up analyses and improve resolution with large datasets.

Model validation and criticism are additional areas in statistical sea-level research that can be greatly improved with standardization of both data compilation and the tools used to evaluate models.  Statistical methods should be validated using cross-validation techniques, bootstrapping, or simulations to precisely determine whether a model achieves what it is designed to. Having a standard set of tools to evaluate the replicability and interrogation of structural assumptions is a clear area for growth in the sea-level community to improve the reproducibility of statistical analysis.


%

\section{Glossary}
\small
\begin{longtable}{p{.25\textwidth}p{.75\textwidth}}
 \caption{Definitions of relevant terms}\\
\multicolumn{1}{l}{\textbf{Term}} & \multicolumn{1}{l}{\textbf{Meaning}} \\ \midrule
\emph{ad hoc}&analysis methods constructed without an underlying statistical theory	\\ \midrule
\emph{analysis choices	}&	decisions in how to implement a specific model structure (e.g., least-squares, likelihood maximization, empirical Bayesian analysis, fully Bayesian analysis (MCMC), Kalman Smoother, ad hoc, EIV, NIGP)\\ \midrule
\emph{Bayesian Hierarchical Model (BHM)}&	uses fully Bayesian analysis, which approximates complicated distributions through sampling, usually using MCMC sampling\\ \midrule
\emph{conditional probability}& the distribution of a random quantity, given a particular value of another (unknown) random quantity; expresses uncertainty in hierarchical models \\ \midrule
\emph{conjugate prior}&	prior distribution that comes from the same family of distributions as the likelihood distribution, so as to enable an analytically-tractable solution for the posterior\\ \midrule
\emph{continuous core	}&	near-continuous records from a single core of salt-marsh sediment or a single coral head\\ \midrule
\emph{covariance function	}&	function defining prior beliefs about the relationship between one or more variables in a Gaussian process, as a measure of how much they change together\\ \midrule
\emph{data level}& hierarchical level that characterizes the relationship between the phenomenon to be modeled and the observed data (instrumental and/or proxy) and incorporates measurement, inferential, and dating uncertainty \\ \midrule
\emph{Empirical Hierarchical Model (EHM)	}&	uses empirical Bayesian analysis, which estimates parameters with point estimates, usually by maximizing their likelihoods, as opposed to a BHM, which samples the prior distributions on parameters\\ \midrule
\emph{errors-in-variable (EIV)	}&	framework that accounts for the measurement uncertainty in the independent variables by assuming that uncertainties in both variables are independent of one another\\ \midrule
\emph{errors-in-variables integrated Gaussian process (EIV-IGP)	}&	model implemented on time series data, modeling the rate of RSL change, deriving the RSL curve from the rate curve; incorporates uncertainty in the independent (time) variable and the dependent variable through EIV framework\\ \midrule
\emph{empirical orthogonal function (EOF)}&	regression technique used to find the dominant spatial patterns in a dataset; when analyzing sea level, used to find the dominant patterns in sea surface height (SSH) from satellite altimetry measurements and apply to tide gauges in order to estimate GMSL change  \\ \midrule
\emph{error	}&	the difference between a measurement and the true value, for a particular data point; one can model the error as a random draw from an uncertainty distribution\\ \midrule
\emph{empirical spatio-temporal Gaussian process  (EST-GP)	}& model with Gaussian process priors, which incorporates spatial and temporal covariance functions to produce the fields of RSL and rate of RSL change as maps; solved using an empirical methodology that maximizes the likelihood of the model conditional upon the parameters of the prior\\ \midrule
\emph{empirical temporal Gaussian process (ET-GP)}&	model using Gaussian process priors, which is independent in space (no spatial component) and solved using an empirical methodology that maximizes the likelihood of the model conditional upon the parameters of the prior\\ \midrule
\emph{Gaussian process (GP)	}&	a generalization of the multi-variate Gaussian distribution to continuous time (and space), which is fully defined by its mean function and covariance function; GP regression provides an analytically-tractable solution when adopting the assumption of normality for all distributions\\ \midrule
\emph{hyperparameter 	}&	parameter of a prior distribution\\ \midrule
\emph{hyperprior 	}&	prior distribution on a hyperparameter\\ \midrule
\emph{inferential uncertainty}&	the quantified dispersion that arises from the data-generation process from true RSL to the creation of a RSL proxy \\ \midrule
\emph{Kalman Smoother (KS)}&	iterative method that comprises a forward filtering pass and a backward smoother pass; used in a multi-model implementation to compute posterior estimates of GMSL and spatio-temporal RSL fields, conditioning prior estimates from physical models of several processes on observations  \\ \midrule
\emph{latent 	}&	unobserved or hidden (e.g., the true values of RSL)\\ \midrule
\emph{likelihood 	}&	the probability of observing the data as described by the fitted model; also known as the sampling or data distribution; a conditional distribution that is a function of unknown parameters for observed data and incorporates the form of uncertainty in the data (e.g., measurement and/or inferential) \\ \midrule
\emph{	marginal distribution	}&	unconditional probability distribution of a random quantity, found by integrating over all values of the conditional distribution in Bayesian analyses\\ \midrule
\emph{Markov Chain Monte Carlo (MCMC)}&	techniques used to generate random variables, perform complicated calculations, and simulate complicated distributions through sampling in Bayesian hierarchical models (common algorithms include Gibbs sampling, Metropolis-Hastings, Metropolis within Gibbs, importance sampling)\\ \midrule
\emph{modeling choices	}&	decisions that define the relationships in a model, usually at the process level; in sea-level analysis, the relationship between time, space and RSL (e.g., linear, polynomial, change-point, GP (integrated), incorporation of physical models)\\ \midrule
\emph{noisy-input Gaussian process (NIGP)}&	a method for incorporating uncertainty in the independent variable within a Gaussian process model; using a Taylor expansion about each input point to recast input noise as output noise proportional to the squared gradient of the GP posterior mean \citep{McHutchon2011Noisy}; in sea-level analysis, geochronological uncertainty is recast as proportional uncertainty in RSL \\ \midrule
\emph{noise	}&	error; statistical noise refers to unexplained variation or randomness\\ \midrule
\emph{noisy data}&	error-prone data that have been corrupted by known or unknown processes\\ \midrule
\emph{non-parametric}&	not involving any assumptions as to the functional form \\ \midrule
\emph{posterior distribution	}& the probability distribution of an unknown quantity, conditional on observed data; in sea-level analysis, estimates (for example) the true RSL time series or field of RSL with uncertainties, given proxy or instrumental data\\ \midrule
\emph{prior distribution}&	the information about an uncertain parameter or process that is combined with the probability distribution of new data to yield the posterior distribution; can be subjective, based on a priori beliefs, or uninformative, which minimizes the impact on inference\\ \midrule
\emph{process level	}&	hierarchical level at which the phenomenon of interest is modeled and in some cases, decomposed; includes process variability\\ \midrule
\emph{residuals	}&	the difference between an observed and a modeled or predicted value\\ \midrule
\emph{sea-level index point (SLIP)	}&	discrete proxy data that constrain RSL at a single point in time and space\\ \midrule
\emph{uncertainty	}&	parameter characterizing the range of values within which a measured value can be said to lie with a specified probability\\ \midrule
\emph{white noise 	}&	serially uncorrelated random variation (zero mean and finite variance)\\ 
\\ \bottomrule
  \label{tab:Glossary}%
\end{longtable}

\section{Acknowledgements:}
This work was supported by National Science Foundation grants OCE-1458904 (ELA, REK, and BPH), OCE-1458903 (SE), OCE-1458921 (ACK), OCE-1702587 (ELA and REK), and is a contribution to IGCP Project 639, INQUA Project CMP1601P ``HOLSEA'' and PALSEA3.

\beginsupplement
\section{Supplemental Information: covariance functions}
\label{sup:cov}

Spatio-temporal covariance functions define a correlation (shared information) through time and space, which typically decays as the time and space differences increase \citep{Rasmussen2006GP}. Some frequently used covariance functions for modeling RSL using GPs include dot-product (e.g., \citealp{Khan2017Carib}), powered-exponential (e.g., \citealp{Cahill2015modeling}), rational quadratic (e.g., \citealp{Kopp2013grl,Hay2015probabilistic}), and Mat\'ern (e.g., \citealp{Hay2015probabilistic,Khan2015Holocene,Kopp2016PNAS,Khan2017Carib}) function.  The choice of prior covariance function(s) characterize stationarity, isotropy, smoothness, and periodicity in Gaussian processes (\ref{SS:IGP}).  For a full treatment of covariance functions, see (\cite{Rasmussen2006GP}, Chapter 4). Each covariance function has distinct characteristics and requires different parameters.  For example, a dot-product covariance function ($K(t_1,t_2)\propto t_1\cdot t_2$) produces a linear trend, which would be appropriate to model GIA over centennial scales.  The powered-exponential covariance function (see equation \ref{eq:pe}) and the Mat\'ern (see equation \ref{eq:Mat}) family of functions are highly generalizable and allow specification of the degree of differentiability (and therefore smoothness), while having a small number of parameters:

\begin{equation}
K(t_i,t_j)=\nu^2 \rho^{|t_i-t_j|^{\kappa}}
\label{eq:pe}
\end{equation}
\noindent where $\rho \in(0,1)$ is the correlation parameter and $\kappa \in(0,2]$ is the smoothness parameter. The IGP employed in \cite{Cahill2015modeling} placed a zero-mean GP prior, with a powered-exponential covariance function on the rate process $f'(t)$. 

A squared-exponential covariance function ($K(t_1,t_2)\propto e^{-\frac{(t_2-t_1)^2}{\theta}}$) is a powered-exponential with a smoothness parameter of two. As the smoothness parameter decreases, the function becomes more rough. The squared-exponential is infinitely differentiable, and is thus very smooth.  Therefore the squared-exponential function would be inappropriate for tectonics, a rough process (except in many passive margin settings), since it would not adequately capture the abrupt changes.  The Mat\'ern is defined as
 
\begin{equation}
\label{eq:Mat}
k_{Matern}(r)=\frac{2^{1-\nu}}{\Gamma{\nu}}\left(\frac{\sqrt{2\nu} r}{l}\right)^{\nu}K_{\nu}\left(\frac{\sqrt{2\nu} r}{l}\right),
\end{equation}
where $r$ is the difference in time or space, $\nu$ is a positive smoothness parameter, $l$ is a positive characteristic length-scale parameter, and $K_{\nu}$ is a modified Bessel function.  When $\nu$ is a half-integer, the covariance function is the product of an exponential and a polynomial, and it is simpler. For example,

\begin{equation}
\label{eq:Mat1}
k_{\nu=1/2}(r)=\left(\frac{r}{l}\right)exp\left(-\frac{r}{l}\right),
\end{equation}
\begin{equation}
\label{eq:Mat3}
k_{\nu=3/2}(r)=\left(1+\frac{\sqrt{3} r}{l}\right)exp\left(-\frac{\sqrt{3} r}{l}\right),
\end{equation}
\begin{equation}
\label{eq:Mat5}
k_{\nu=5/2}(r)=\left(1+\frac{\sqrt{5} r + \frac{5r^2}{3l^2}}{l}\right)exp\left(-\frac{\sqrt{5} r}{l}\right),
\end{equation}
There is some trade-off between the Mat\'ern exponent values (and thus the degree of GP differentiability) and the characteristic length scale parameter: the selection of a lower exponent (which creates a less smooth function) is somewhat comparable, when it has a longer length scale, to a covariance function with a greater exponent and a shorter length scale \citep{Hay2015probabilistic}. 

The sum of several covariance functions can be used to model the RSL field, with each term separated by spatial or temporal scales.  While it may not be possible to explicitly distinguished between sea-level processes through the characteristic scale hyperparameters alone, information from physical models can be incorporated into the covariance structure of GPs.  For example, in the GP model from \cite{Hay2015probabilistic}, the melt component $M(\bm{x},t)$ was the sum of individual ice sheet or mountain glacier covariances, where each had a linear term and a rational quadratic term, both of which were dependent in time.  The sum of the two terms was multiplied by a spatial weighting $B^M$, which applied the sea-level fingerprint associated with the melt for each land-based ice source. The covariance function of the GP prior for the melt was:
\begin{equation}
\sum_{j=1}^{n}{M(\bm{x},t)} = \sum_{j=1}^n \left( m_a\cdot\Delta t_{q,p} + c \left( 1+ \frac{\Delta t^2_{q,p}}{2 \alpha \tau_M^2} \right)^{-\alpha} \right) B^M
\end{equation}
\noindent where $j$ represents each ice sheet or glacier, $t_q$ and $t_p$ represent the time at the $q$th and $p$th time step, $\Delta t_{q,p}$ represents the time difference between the steps, and $m$, $c$, $\alpha$, and $\tau_M$ are hyperparameters that defined the prior standard deviation of the linear rate, the prior standard deviation on non-linear variability and the roughness and characteristic timescale of non-linear variability.  These parameters were estimated by maximizing the likelihood of published reconstructions of the time series of glacier and ice sheet estimates.

\section{Supplemental Information: Analysis Choice Details}
\subsection{Virtual station} \label{SI:RA}
The virtual station approach separates the ocean into pre-defined oceanic regions or basins. The methodology averages two monthly-mean tide gauges together, creating a virtual station located at the halfway point between the two original stations.  This averaging is repeated until one virtual station exists in each region. The global average is then computed by averaging all of the virtual stations.  \citet{Jevrejeva2006,jevrejeva2008recentaccel} also removed 2-30 year variability using a method based on Monte Carlo Singular Spectrum Analysis.

In an extension of the original virtual station technique of \cite{Jevrejeva2006,jevrejeva2008recentaccel}, \cite{Dangendorf2017GMSL}
computed regional mean sea-level rates from subsets of tide gauges after first correcting for processes that affect RSL and SSH, such as GIA, vertical land motion, and geoid changes due to glacier melting. The regional averages were then combined by weighting each region by the area of the ocean it represents.

\subsection{Kalman smoother}
\label{Sup:KS}

In the Kalman smoother, $\mathbf{y}_k$, which are observations of RSL at each time step taken from a global network of tide gauge sites, are modeled as:

\begin{equation}
\label{eq:KS3}
\mathbf{y}_k = \mathbf{Hx}_k + \mathbf{v}_k
\end{equation}
\noindent where the observation matrix, $\mathbf{H}$, maps the state vector into the observation space, and the measurement noise, $\mathbf{v}_k$, is assumed to have a mean of zero with covariance $\mathbf{R}$. 

Equation \ref{eq:KSProc} can be reframed in KS terminology, where the spatial sea-level field $\bm{f}_k$ is a vector of local RSLs at time step $k$ and locations of interest, and $\bm{\beta}_k$ is a vector that contains the melt rates $M_j$ of 18 mountain glaciers, 3 ice sheets, and a globally uniform term, $g(t)$. At each time step, the filter constructs a prior estimate of the state vector, $\mathbf{x}_k$, defined as:

\begin{equation}
\mathbf{x}_k = \left[ \begin{array} {c}
\bm{f}_k \\ \bm{\beta}_k   \end{array} \right]
 = \bm{\Phi} \mathbf{x}_{k-1} + \mathbf{Bu}_k + \mathbf{w}
\end{equation}

\noindent where $\mathbf{\Phi}$ is the state transition matrix, $\mathbf{u}_k$ is the input control parameter, $\mathbf{B}$ maps the input control parameter into the state vector, and $\mathbf{w}$ is the zero-mean process noise with covariance $\mathbf{Q}$.  The normalized sea-level fingerprints ($FP_j$ from equation \ref{eq:KSProc}), which connect local RSL to the global average melt rates being estimated, are contained in $\mathbf{\Phi}$, and $\mathbf{u}_k$ includes the rates of local sea-level change controlled by both GIA and ocean dynamics (see \citealp{Hay2013GSL} for an explicit description of each matrix). 

The Kalman filter consists of two main steps: the time update step and the measurement update step. In the time update step, the filter constructs a prior estimate of the state vector, $\mathbf{\hat{x}}_k^{-}$, and its covariance, $\mathbf{P}_k^{-}$, at time step $k$ conditioned upon the state vector at time $k-1$.  The superscript 
minus sign, $^-$, indicates that the estimate is computed in the time update stage (prediction) and represents the prior estimate of the states. 

\begin{align}
\label{eq:KS4}
\mathbf{\hat{x}}_k^{-} &= \mathbf{\Phi x}_{k-1} + \mathbf{Bu}_k\\
\label{eq:KS5}
\mathbf{P}_k^{-} &= \mathbf{\Phi P}_{k-1} \mathbf{\Phi}^T + \mathbf{Q}
\end{align}
The time update step, described by equations \ref{eq:KS4} and \ref{eq:KS5}, contains all the process-based physical models of the drivers of sea-level change.

In the measurement update step, the prior estimates, $\mathbf{\hat{x}}_k^{-}$ and $\mathbf{P}_k^{-}$, are conditioned upon the available observations $\mathbf{z}_k$ at time $k$.  The goal is to find the optimal estimate of the state vector, $\mathbf{\hat{x}}_k$, and covariance, $\mathbf{P}_k$, that combines the prior estimates with the observations: 

\begin{align}
\label{eq:KS6}
\mathbf{\hat{x}}_k &= \mathbf{\hat{x}}_k^{-} + K_k(\mathbf{z}_k-\mathbf{H\hat{x}}_k^{-})\\
\label{eq:KS7}
\mathbf{P}_k &= (\mathbf{I}-\mathbf{K}_k\mathbf{H})\mathbf{P}_k^{-}
\end{align}
Here $\mathbf{K}_k$ is the Kalman gain matrix defined as
\begin{align}
\label{eq:KS8}
\mathbf{K}_k &= \mathbf{P}_k^{-}\mathbf{H}^T(\mathbf{HP}_k^{-}\mathbf{H}^T + \mathbf{R})^{-1}
\end{align}
The prediction and measurement steps (equations \ref{eq:KS4}, \ref{eq:KS5}, \ref{eq:KS6}, and \ref{eq:KS7}) are recursively computed through time until all observations have been assimilated \citep{Kalman1960KS}.

Once the forward pass is complete, the Kalman filter is run backwards in time and a weighted combination of the forward and backwards passes is computed. This three-pass-filter \citep{Gelb1974Interp} ensures that in every year the optimal estimate of the state vector and its covariance includes all observations over the analysis time window.

The final component of the multi-model Kalman smoother implemented by \cite{Hay2015probabilistic} is the multi-model step. In this step, the likelihood of obtaining the observations, given the model, is computed.  These probabilities are then used to compute a weighted sum of the Kalman smoother estimates \citep{Blom1988Ieee}. 

In \cite{Hay2015probabilistic}, the melt rates $M_j(\bm{x},t)$ along with their associated covariance, are summed to produce an estimate of GMSL over the 20th century. Unmodeled local processes, such as tectonics and groundwater withdrawal, are not explicitly modeled in the Kalman smoother and are therefore not mapped into GMSL. Instead, these unmodeled effects are captured by the process noise term. While the unmodeled local processes are not included in GMSL, they are present in the Kalman-smoother reconstructed tide gauge and global sea-level time series. An alternative approach for reconstructing the local sea-level time series is to use the Kalman smoother posterior estimates of the melt rates (with the normalized fingerprints), uniform sea-level change, GIA, and ocean dynamics. The site-specific components of each of these processes can be summed together to reconstruct the local sea-level field. This field will not contain the local processes that are observed in the data since they are not mapped into the individual components estimated in the smoother. It will, therefore, be an inherently smoother reconstruction and will have lower uncertainties than the field estimated within the Kalman smoother \citep{Hay2017FP}.

\section{Supplemental Information: Details of implementations}
\label{SI:Imp}

\subsection{Time series models and implementations}
\label{SI:TS_Mods}
\noindent \textbf{Temporally linear model:}  Using a simple, temporally linear model, we applied ordinary (OLS) and general (GLS) least-squares analyses to the continuous core records.  The OLS analysis was conducted on the mean RSL and median age for each continuous core record (i.e., excluding vertical and temporal uncertainties).  The GLS analysis included vertical (RSL) uncertainty (Figure \ref{fig:comp}).  The estimated OLS and GLS parameters are similar for the two sites (Figure \ref{fig:param}, table a).

\noindent \textbf{Change-point regression:} We implemented a linear change-point model with fully Bayesian analysis, and incorporated temporal uncertainty within an errors-in-variable framework. The NJ record was best fit by a model with three change points, whereas the NC record was best fit with two change points (Figure \ref{fig:param}, table b). 
\label{SS:EHMGPImp}

\noindent \textbf{Empirical temporal GP model (ET-GP):} We implemented a temporal-only model with GP priors using empirical Bayesian analysis with the following process model:
\begin{equation}
f(t)=l(t)+m(t)+w(t)
\end{equation}

\noindent where $l(t)$ and $m(t)$ are low- and medium-frequency terms, respectively, and $w(t)$ is high-frequency variability, interpreted as white noise.  The $l(t)$ and $m(t)$ terms were each assigned zero-mean GP priors with Mat\'ern($3/2$) covariance functions yield curves that are once differentiable everywhere (see Section \ref{sup:cov} for more details on the smoothness and other characteristics of various covariance functions).  Using empirical Bayesian analysis, the optimal point estimates of the hyperparameters varied for the two independent models (Figure \ref{fig:param}, table d).
\label{SS:IGPImp}

\noindent\textbf{Bayesian EIV-IGP:} We implemented the EIV-IGP model with fully Bayesian analysis (Section \ref{SS:IGP}), where the posterior distributions on the hyperparameters differed for the two models (Figure \ref{fig:param}, table c).

\subsection{Spatio-temporal models and implementations}	\label{SI:ST}
\label{SS:EST-GPR}
\noindent \textbf{Empirical spatio-temporal GP model (EST-GP):}
We implemented an empirical spatio-temporal GP model (EST-GP) using discrete index points and continuous core records with the process model in equation \ref{eq:pure_stat} with an addition white noise term, $w(t)$.  In this implementation, $g(t)$, $r(\bm{x},t)$, and $m(\bm{x},t)$ are common, regional, and local terms, respectively, each with zero-mean GP priors with Mat\'ern($3/2$) covariance functions.  

\noindent\textbf{Bayesian EIV-IGP for single sites:} As in Section \ref{SS:IGPImp}, we implemented the Bayesian EIV-IGP (Section \ref{SS:IGP}), which does not include a spatial component, in order to provide a comparison to the spatio-temporal model.  The input data for New Jersey and North Carolina were equivalent to the EST-GP, including index points and continuous core records from each location.

\subsection{Instrumental models and implementations}
\label{SI:instr}
\noindent \textbf{Empirical spatio-temporal GP model (EST-GP):} Using a regional subset of tide gauges from the U.S. Atlantic coast (from the same geographic range as the proxy data in Section \ref{SS:EST-GPR}), we implemented an empirical spatio-temporal GP model (EST-GP).  Results include estimates of the spatio-temporal fields of RSL and its rates change with uncertainties along the U.S. Atlantic coast and estimates at two specific sites (Figure \ref{fig:KS_GPR}).  Although this technique can produce estimates of GMSL (as in \citealp{Hay2015probabilistic}), this implementation does not include global results, because it is implemented on a regional subset of data.

\noindent\textbf{KS implementation}
Using a subset of global tide gauges (see \citealp{Hay2015probabilistic} for further information on the tide gauges used), we implemented the process model in equation \ref{eq:KSProc} with a multi-model KS (Section \ref{SS:KST}).  The KS estimates GMSL and the spatio-temporal fields of RSL and rates of RSL change along the U.S. Atlantic coast and at specific sites (Figure \ref{fig:KS_GPR}).

\subsection{Supplemental model predictions and hyperparameter results}
\label{SI:other}

\begin{figure}[ht]
\centering\includegraphics[width=0.9\linewidth]{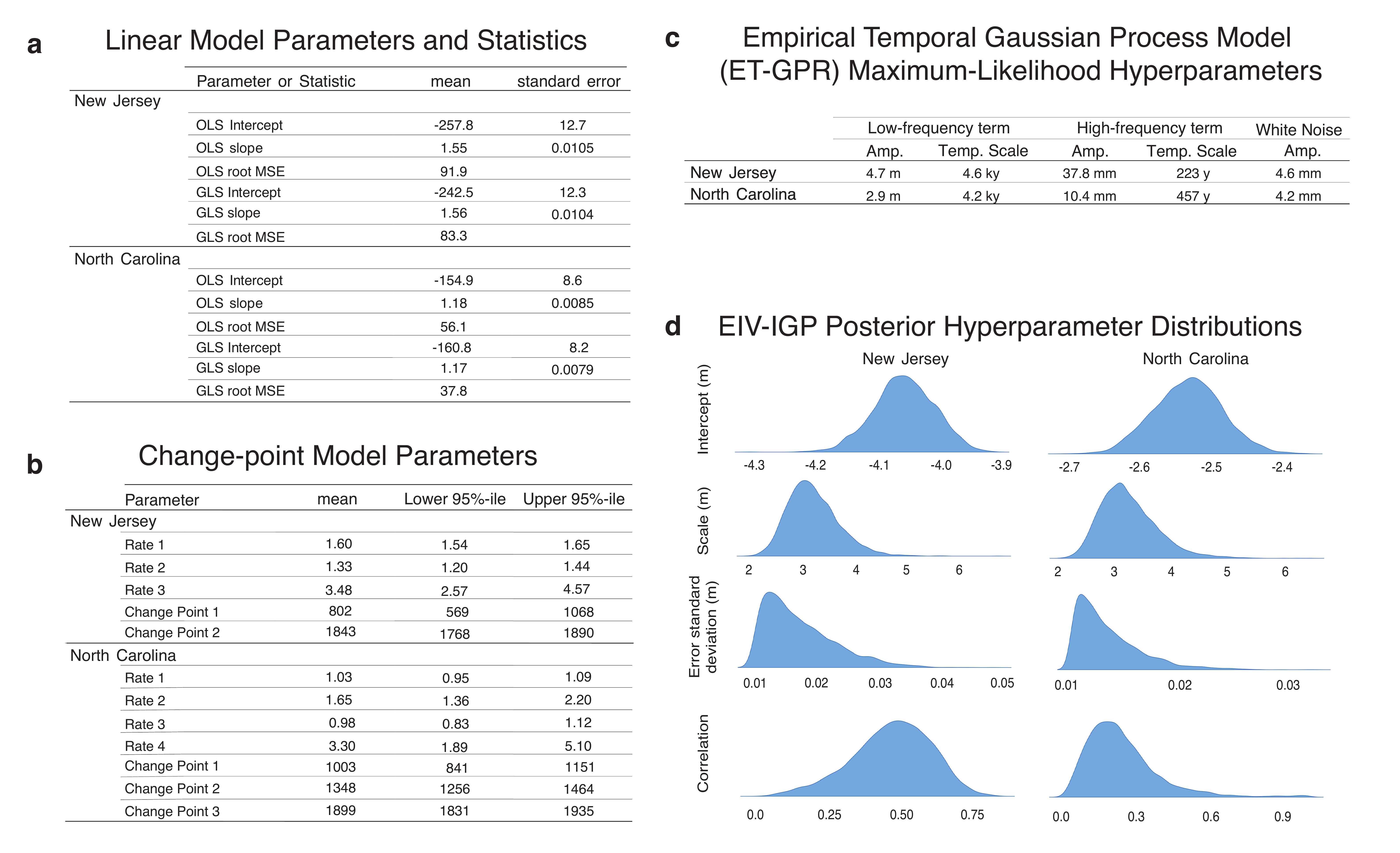}
\caption{(a) Table showing linear model parameters and statistics, including intercept and slope with their standard errors and root mean squared error, for two versions of least-squares linear implementation (OLS and GLS). (b) Table showing change points fit to continuous core records at New Jersey and North Carolina, where the mean and 95\% credible intervals are shown for the rates and change points. (c) Bayesian EIV-IGP posterior distributions of hyperparameters for each of the independent models at New Jersey and North Carolina.  The intercept is $\alpha$ in equation \ref{eq:IGP}, the scale is $\nu$ in equation \ref{eq:pe}, the standard deviation of the measurement uncertainty is $\sigma$ as in equation \ref{eq:TSData}, and the correlation is $\rho$ from equation \ref{eq:pe}.  (d) Table of maximum-likelihood hyperparameters for independent empirical-temporal GP models in New Jersey and North Carolina, including prior amplitude and temporal-scale parameters of the low-frequency and high-frequency terms, the white noise amplitude.}
\label{fig:param}
\end{figure}

The hyperparameters for each of the time-series implementations are summarized in Figure \ref{fig:param}.

\begin{table}[htbp]
  \centering
  \caption{RSL predictions (m) with 95\% CI below from EST-GPR}
  \resizebox{\textwidth}{!}{%
    \begin{tabular}{ccccccccccccccccccccccccccccccccc}
    Site  & Lat   & Lon   & \multicolumn{2}{c}{14 ka} & \multicolumn{2}{c}{13 ka} & \multicolumn{2}{c}{12 ka} & \multicolumn{2}{c}{11 ka} & \multicolumn{2}{c}{10 ka} & \multicolumn{2}{c}{9 ka} & \multicolumn{2}{c}{8 ka} & \multicolumn{2}{c}{7 ka} & \multicolumn{2}{c}{6 ka} & \multicolumn{2}{c}{5 ka} & \multicolumn{2}{c}{4 ka} & \multicolumn{2}{c}{3 ka} & \multicolumn{2}{c}{2 ka} & \multicolumn{2}{c}{1 ka} & \multicolumn{2}{c}{0 ka} \\
\midrule
 \multirow{2}[0]{*}{New Jersey} & \multirow{2}[0]{*}{39.09} & \multirow{2}[0]{*}{284.77} & \multicolumn{2}{c}{-59.0} & \multicolumn{2}{c}{-51.2} & \multicolumn{2}{c}{-43.0} & \multicolumn{2}{c}{-35.1} & \multicolumn{2}{c}{-28.1} & \multicolumn{2}{c}{-22.4} & \multicolumn{2}{c}{-18.2} & \multicolumn{2}{c}{-14.6} & \multicolumn{2}{c}{-11.4} & \multicolumn{2}{c}{-9.1} & \multicolumn{2}{c}{-7.0} & \multicolumn{2}{c}{-4.9} & \multicolumn{2}{c}{-3.4} & \multicolumn{2}{c}{-1.7} & \multicolumn{2}{c}{-0.1} \\
          &       &       & -63.9 & -54.1 & -54.6 & -47.8 & -45.3 & -40.7 & -36.7 & -33.5 & -29.2 & -27.0 & -23.1 & -21.6 & -18.7 & -17.7 & -15.1 & -14.1 & -11.8 & -11.0 & -9.4  & -8.8  & -7.3  & -6.7  & -5.2  & -4.7  & -3.6  & -3.2  & -1.9  & -1.6  & -0.1  & 0.0 \\         
\multirow{2}[0]{*}{North Carolina} & \multirow{2}[0]{*}{34.98} & \multirow{2}[0]{*}{283.8} & \multicolumn{2}{c}{-53.1} & \multicolumn{2}{c}{-45.2} & \multicolumn{2}{c}{-36.9} & \multicolumn{2}{c}{-29.0} & \multicolumn{2}{c}{-22.0} & \multicolumn{2}{c}{-16.4} & \multicolumn{2}{c}{-12.5} & \multicolumn{2}{c}{-9.4} & \multicolumn{2}{c}{-6.9} & \multicolumn{2}{c}{-5.4} & \multicolumn{2}{c}{-4.2} & \multicolumn{2}{c}{-2.9} & \multicolumn{2}{c}{-2.0} & \multicolumn{2}{c}{-1.3} & \multicolumn{2}{c}{0.1} \\
          &       &       & -57.5 & -48.7 & -48.0 & -42.4 & -38.6 & -35.2 & -30.2 & -27.7 & -23.1 & -20.8 & -17.5 & -15.3 & -13.5 & -11.4 & -10.3 & -8.5  & -7.6  & -6.2  & -5.9  & -4.8  & -4.6  & -3.7  & -3.2  & -2.5  & -2.2  & -1.8  & -1.4  & -1.1  & -0.1  & 0.2 \\  
   \multirow{2}[0]{*}{Inner Delaware} & \multirow{2}[0]{*}{38.75} & \multirow{2}[0]{*}{284.88} & \multicolumn{2}{c}{-58.6} & \multicolumn{2}{c}{-50.8} & \multicolumn{2}{c}{-42.5} & \multicolumn{2}{c}{-34.7} & \multicolumn{2}{c}{-27.7} & \multicolumn{2}{c}{-22.0} & \multicolumn{2}{c}{-17.8} & \multicolumn{2}{c}{-14.3} & \multicolumn{2}{c}{-11.2} & \multicolumn{2}{c}{-8.9} & \multicolumn{2}{c}{-6.9} & \multicolumn{2}{c}{-4.9} & \multicolumn{2}{c}{-3.4} & \multicolumn{2}{c}{-1.7} & \multicolumn{2}{c}{-0.1} \\
          &       &       & -57.5 & -48.7 & -48.0 & -42.4 & -38.6 & -35.2 & -30.2 & -27.7 & -23.1 & -20.8 & -17.5 & -15.3 & -13.5 & -11.4 & -10.3 & -8.5  & -7.6  & -6.2  & -5.9  & -4.8  & -4.6  & -3.7  & -3.2  & -2.5  & -2.2  & -1.8  & -1.4  & -1.1  & -0.1  & 0.2 \\
\end{tabular}}%
  \label{tab:table2}%
\end{table}%

\begin{table}[htbp]
  \centering
  \caption{Predicted rates (m/ky) averaged over 1000 year periods from EST-GP with 95\% CI below}
  \resizebox{\textwidth}{!}{%
    \begin{tabular}{ccccccccccccccccccccccccccccc}
          & \multicolumn{2}{c}{13-14 ka} & \multicolumn{2}{c}{12-13 ka} & \multicolumn{2}{c}{11-12 ka} & \multicolumn{2}{c}{10-11 ka} & \multicolumn{2}{c}{9-10 ka} & \multicolumn{2}{c}{8-9 ka} & \multicolumn{2}{c}{7-8 ka} & \multicolumn{2}{c}{6-7 ka} & \multicolumn{2}{c}{5-6 ka} & \multicolumn{2}{c}{4-5 ka} & \multicolumn{2}{c}{3-4 ka} & \multicolumn{2}{c}{2-3 ka} & \multicolumn{2}{c}{1-2 ka} & \multicolumn{2}{c}{0-1 ka}\\
           \midrule 
   \multirow{2}[0]{*}{New Jersey} & \multicolumn{2}{c}{7.8} & \multicolumn{2}{c}{8.2} & \multicolumn{2}{c}{7.9} & \multicolumn{2}{c}{7.0} & \multicolumn{2}{c}{5.7} & \multicolumn{2}{c}{4.2} & \multicolumn{2}{c}{3.6} & \multicolumn{2}{c}{3.2} & \multicolumn{2}{c}{2.3} & \multicolumn{2}{c}{2.1} & \multicolumn{2}{c}{2.1} & \multicolumn{2}{c}{1.5} & \multicolumn{2}{c}{1.7} & \multicolumn{2}{c}{1.7} \\
          & 5.9   & 9.7   & 6.7   & 9.7   & 6.8   & 9.0   & 6.2   & 7.8   & 5.1   & 6.4   & 3.7   & 4.7   & 3.2   & 4.0   & 2.8   & 3.6   & 2.0   & 2.7   & 1.8   & 2.4   & 1.8   & 2.4   & 1.2   & 1.8   & 1.5   & 1.9   & 1.5   & 1.8 \\
    \multirow{2}[0]{*}{North Carolina} & \multicolumn{2}{c}{8.0} & \multicolumn{2}{c}{8.3} & \multicolumn{2}{c}{7.9} & \multicolumn{2}{c}{7.0} & \multicolumn{2}{c}{5.5} & \multicolumn{2}{c}{4.0} & \multicolumn{2}{c}{3.1} & \multicolumn{2}{c}{2.5} & \multicolumn{2}{c}{1.5} & \multicolumn{2}{c}{1.2} & \multicolumn{2}{c}{1.3} & \multicolumn{2}{c}{0.9} & \multicolumn{2}{c}{0.7} & \multicolumn{2}{c}{1.3} \\
 \centering
 & 6.07 & 9.8   & 6.8   & 9.8   & 6.9   & 9.0   & 6.3   & 7.7   & 5.0   & 6.1   & 3.5   & 4.5   & 2.6   & 3.6   & 2.0   & 2.9   & 1.1   & 2.0   & 0.8   & 1.6   & 0.9   & 1.7   & 0.6   & 1.2   & 0.5   & 1.0   & 1.2   & 1.5 \\
    \multirow{2}[0]{*}{Inner Delaware} & \multicolumn{2}{c}{7.8} & \multicolumn{2}{c}{8.2} & \multicolumn{2}{c}{7.9} & \multicolumn{2}{c}{7.0} & \multicolumn{2}{c}{5.7} & \multicolumn{2}{c}{4.2} & \multicolumn{2}{c}{3.5} & \multicolumn{2}{c}{3.1} & \multicolumn{2}{c}{2.2} & \multicolumn{2}{c}{2.0} & \multicolumn{2}{c}{2.0} & \multicolumn{2}{c}{1.5} & \multicolumn{2}{c}{1.7} & \multicolumn{2}{c}{1.6} \\
          & 5.9   & 9.7   & 6.7   & 9.7   & 6.8   & 9.0   & 6.2   & 7.8   & 5.0   & 6.3   & 3.6   & 4.7   & 3.1   & 4.0   & 2.7   & 3.5   & 1.9   & 2.6   & 1.7   & 2.4   & 1.7   & 2.3   & 1.2   & 1.7   & 1.5   & 1.9   & 1.5   & 1.7 \\
    \end{tabular}}%
  \label{tab:table3}%
\end{table}%
\section{Supplemental Information: Data}
\label{SI:Dat}
The data used in the spatio-temporal models can be found in the following files, and the data used in the independent temporal models is a subset of the following file, as described in Section \ref{S:Implement}:\\ 
Holocene\_Data.csv\\

\bibliographystyle{model2-names}
\bibliography{Sources}

\end{document}